\documentclass[12pt]{article}
\usepackage[a4paper,margin=1in]{geometry}
\usepackage[titletoc,title]{appendix}
\usepackage{changepage}
\usepackage[utf8x]{inputenc}
\usepackage{textcomp,marvosym}
\usepackage{fixltx2e}
\usepackage{amsmath,amssymb}
\usepackage{cite}
\usepackage{nameref,hyperref}
\usepackage{authblk}
\usepackage{algorithm}
\usepackage{algpseudocode}
\usepackage{dsfont}
\usepackage{xcolor}
\usepackage{enumitem}
\usepackage{caption,setspace}
\usepackage{subcaption}
\usepackage{pdflscape}
\usepackage{graphicx}
\usepackage{placeins}
\usepackage{extarrows}
\usepackage{natbib}
\newcommand{\dydx}[2]{\frac{\text{d} #1}{\text{d} #2}}
\newcommand{\ddydx}[2]{\frac{\text{d}^2 #1}{\text{d} {#2}^2}}
\newcommand{\pdydx}[2]{\frac{\partial #1}{\partial #2}}
\newcommand{\pddydx}[2]{\frac{\partial^2 #1}{\partial {#2}^2}}

\renewcommand{\eqref}[1]{Equation~(\ref{#1})}
\newcommand{\Prob}[1]{\mathbb{P}(#1)}
\newcommand{\CondProb}[2]{\mathbb{P}(#1 \mid #2)}
\newcommand{\CondE}[2]{\mathbb{E}\left[#1 \mid #2\right]}

\newcommand{\PDF}[1]{p(#1)}
\newcommand{\CondPDF}[2]{p(#1 \mid #2)}
\newcommand{\hatCondPDF}[2]{\hat{p}(#1 \mid #2)}

\newcommand{\like}[2]{\mathcal{L}(#1 ; #2)}
\newcommand{\mclike}[2]{\hat{\mathcal{L}}(#1 ; #2)}
\newcommand{\Kernel}[2]{q(#1 \mid #2)}
\newcommand{\E}[1]{\mathbb{E}\left[#1\right]}
\newcommand{\V}[1]{\text{Var}\left[#1\right]}

\newcommand{\C}[2]{\text{Cov}\left[#1,#2\right]}
\newcommand{\bvec}[1]{\mathbf{#1}}

\newcommand{\simProc}[2]{f(#1 \mid #2)}
\newcommand{\obsProc}[2]{g(#1 \mid #2)}
\newcommand{\paramvec}{\boldsymbol{\theta}}
\newcommand{\param}{\theta}
\newcommand{\paramspace}{\boldsymbol{\Theta}}
\newcommand{\discrep}[2]{\rho( #1, #2)}
\newcommand{\dat}{\mathcal{D}}
\newcommand{\simdat}{\mathcal{D}_s}

\newcommand{\doi}[1]{DOI:\href{https://doi.org/#1}{#1}}

\title{A practical guide to pseudo-marginal methods for computational inference in systems biology}

\author[1]{David~J. Warne\footnote{To whom correspondence should be addressed. E-mail: david.warne@qut.edu.au}}
\author[2]{Ruth~E. Baker}
\author[1]{Matthew~J. Simpson}

\affil[1]{School of Mathematical Sciences, Queensland University of Technology, Brisbane, Queensland 4001, Australia}
\affil[2]{Mathematical Institute, University of Oxford, Oxford, OX2 6GG, United Kingdom}

\begin{document}

\maketitle
\begin{abstract}
For many stochastic models of interest in systems biology, such as those describing biochemical reaction networks, exact quantification of parameter uncertainty through statistical inference is intractable. Likelihood-free computational inference techniques enable parameter inference when the likelihood function for the model is intractable but the generation of many sample paths is feasible through stochastic simulation of the forward problem. The most common likelihood-free method in systems biology is approximate Bayesian computation that accepts parameters that result in low discrepancy between stochastic simulations and measured data. However, it can be difficult to assess how the accuracy of the resulting inferences are affected by the choice of acceptance threshold and discrepancy function. The pseudo-marginal approach is an alternative likelihood-free inference method that utilises a Monte Carlo estimate of the likelihood function. This approach has several advantages, particularly in the context of noisy, partially observed, time-course data typical in biochemical reaction network studies. Specifically, the pseudo-marginal approach facilitates exact inference and uncertainty quantification, and may be efficiently combined with particle filters for low variance, high-accuracy likelihood estimation. In this review, we provide a practical introduction to the pseudo-marginal approach using inference for biochemical reaction networks as a series of case studies. Implementations of key algorithms and examples are provided using the Julia programming language; a high performance, open source programming language for scientific computing (\href{https://github.com/davidwarne/Warne2019_GuideToPseudoMarginal}{https://github.com/davidwarne/Warne2019\_GuideToPseudoMarginal}).          
\end{abstract}


\pagebreak

\section{Introduction}

Stochastic models are routinely used in systems biology to facilitate the interpretation and understanding of experimental observations. In particular,
stochastic models are often more realistic descriptions, compared with their deterministic counterparts, of many biochemical processes that are naturally affected by extrinsic and intrinsic noise~\citep{Kaern2005,Raj2008}, such as the biochemical reaction pathways that regulate gene expression~\citep{Paulsson2000,Tian2006}. Such stochastic models enable the exploration of various biochemical network motifs to explain particular phenomena observed though the use of modern, high resolution experimental techniques~\citep{Sahl2017}. The validation and comparison of theories against observations can be achieved using statistical inference techniques to quantify the uncertainty in unknown parameters and likelihoods of observations under different models.   
Recent reviews by ~\cite{Schnoerr2017} and \cite{Warne2019} highlight the  state-of-the-art in computational techniques for simulation of biochemical networks, analysis of the distribution of future states of the biochemical systems, and computational inference from a Bayesian perspective. Both studies point out that, for realistic biochemical reaction networks, the likelihood function is intractable. As a result, likelihood-free computation inference schemes are essential for practical situations.

In our previous work~\citep{Warne2019}, we provide an accessible discussion of a wide range of algorithms for simulation and inference in the context of biochemical systems and provide example implementations for demonstration purposes. In particular, \cite{Warne2019} highlights the use of approximate Bayesian computation (ABC)~\citep{Sisson2018} for likelihood-free inference of kinetic rate parameters using time-course data. While ABC is a widely applicable and popular likelihood-free approach within the life sciences~\citep{Toni2008}, inferences obtained by this method are, as the name implies, approximations, and the accuracy of these approximations are highly dependent upon choices made by the user~\citep{Sunnaker2013}. 

\begin{figure}[h]
	\centering
	\includegraphics[width=\linewidth]{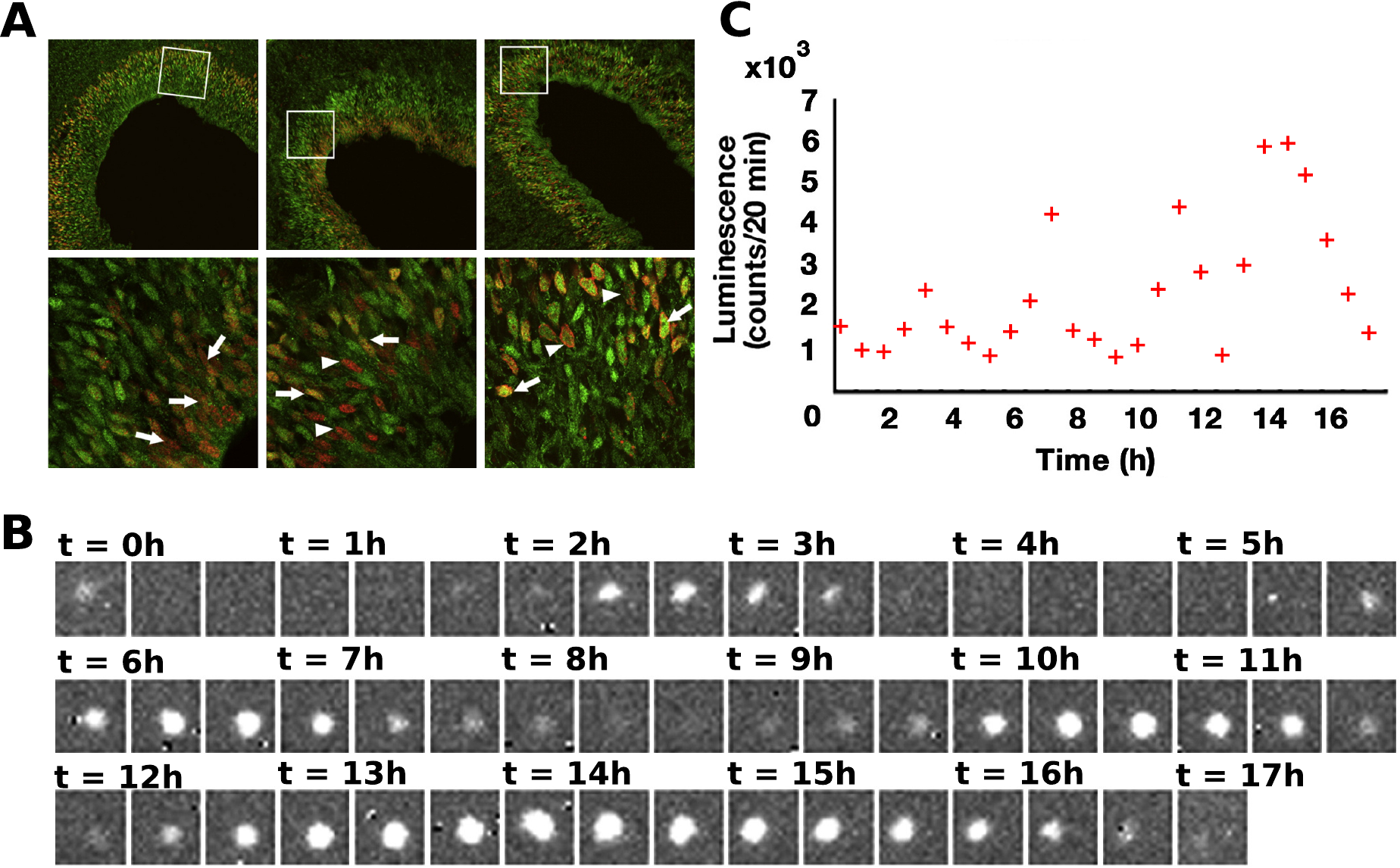}
	\caption{Time-course gene expression data. (A) snapshots of time-lapse microscopy using two fluorescent reporters indicating Hes1 gene expression (green) and BrdU incorporation (red) used to indicate cell-cycle phases. (B) Time-series of green fluorescent reporter for a tracked single-cell over 17 hours in 20 minute intervals. (C) the resulting time-series indicating oscillatory Hes1 gene expression. Panels (A),(B), and (C) are modified with permission from~\cite{Shimojo2008}.}
	\label{fig:data}
\end{figure}

Time-course data describing temporal variations in particular molecular signals within living cells are often obtained using time-lapse optical microscopy with fluorescent reporters (Figure~\ref{fig:data}(A))~\citep{BarJoseph2012,Locke2009,Young2011}. Individual cells are tracked (Figure~\ref{fig:data}(B)) and the luminescence from the reporter is measured over time at discrete intervals (Figure~\ref{fig:data}(C)). These luminescence values are then used to determine concentrations of mRNAs or proteins that may be associated with the expression of a particular gene over time. These data provide information about the complex dynamics of gene regulatory networks that can result in stochastic switching~\citep{Tian2006} or oscillatory behaviour (Figure~\ref{fig:data}(C))~\citep{Elowitz2000,Shimojo2008}. 

Common features of gene expression time-course data include sparsity of temporal observations, relatively few concurrent fluorescent reporters, and noisy observations. Therefore, likelihood-free inference methods are essential to deal with statistical inference~\citep{Toni2008}. However, complex dynamics observed in real gene regulatory networks, such as stochastic oscillations or bi-stability,  can render ABC methods impractical for accurate inferences since most simulations will be rejected, even when the values of model parameters are close to the true values.

Pseudo-marginal methods~\citep{Andrieu2009} are an alternative likelihood-free approach that can provide exact inferences under the prescribed model and are significantly less sensitive to user-defined input. Variants of this approach are particularly well suited for Bayesian inference of nonlinear stochastic models using partially observed time-course data~\citep{Andrieu2010}. This makes the pseudo-marginal method ideal for the study of biochemical systems, however, to-date, few applications of these approaches are present in the systems biology literature~\citep{Golightly2008,Golightly2011}. While \cite{Warne2019} briefly discuss the pseudo-marginal approach, no examples or implementations are provided.

The purpose of this review is to complement \cite{Warne2019} and~\cite{Schnoerr2017} by providing an accessible, didactic guide to pseudo-marginal methods~\citep{Andrieu2009,Andrieu2010,Doucet2015} for the inference of kinetic rate parameters of biochemical reaction network models using the chemical Langevin description. For all of our examples, we provide accessible implementations using the open source, high performance Julia programming language~\citep{Besancon2019,Bezanson2017}\footnote{The Julia code examples and demonstration scripts are available from GitHub \href{https://github.com/davidwarne/Warne2019_GuideToPseudoMarginal}{https://github.com/davidwarne/Warne2019\_GuideToPseudoMarginal}}.

\section{Background}
\label{sec:back}
In this section, we introduce several concepts that are fundamental to understanding how pseudo-marginal methods work and why they are effective. Firstly, we introduce stochastic biochemical reaction networks and how one might model and simulate these systems using stochastic differential equations (SDEs). The Bayesian inference framework is then described along with the essentials of Markov chain Monte Carlo (MCMC) sampling. Lastly, an analytically tractable inference problem is presented, along with Julia code implementations, in order to solidify the concepts, as they are relied upon in subsequent sections.   

\subsection{Stochastic biochemical reaction networks}
\label{sec:intro_sde}
A biochemical reaction network consists of $N$ chemical species, $X_1, X_2, \ldots, X_N$ that interact via a network of $M$ reactions,
\begin{equation}
\sum_{i=1}^N \nu_{i,j}^- X_i \overset{k_j}{\rightarrow} \sum_{i=1}^N \nu_{i,j}^+ X_i, \quad j = 1,2,\ldots, M,
\end{equation}
where $\nu_{i,j}^-$ and $\nu_{i,j}^+$ are, respectively, the number of reactant and product molecules of species $X_i$ involved in the $j$th reaction, and $k_j$ is the kinetic rate parameter for the $j$th reaction. We refer to $\nu_{i,j} = \nu_{i,j}^+ - \nu_{i,j}^-$ as the stoichiometry of species $i$ for reaction $j$. While spatially extended systems can be considered~\citep{Cotter2016,Flegg2015}, we will assume the chemical mixture is spatially uniform for clarity. Under this assumption the law of mass action holds and the probability of the $j$th reaction occurring in the time interval $[t, t + \text{d}t)$ is $a_j(\bvec{X}_t)\text{d}t$, where $\bvec{X}_t = [X_{1,t},X_{2,t},\ldots,X_{N,t}]^\text{T}$ is an $N \times 1$ state vector consisting of the copy numbers for each species at time $t$ and $a_j(\bvec{X}_t)$ is the \emph{propensity function} for reaction $j$~\citep{Gillespie1977,Kurtz1972}. Should a reaction $j$ event occur then the state will update by adding the stoichiometric vector  $\boldsymbol{\nu}_j = [\nu_{1,j},\nu_{2,j},\ldots,\nu_{N,j}]^\text{T}$ to the current system state. Example implementations for generating a range of common biochemical reaction network models is provided in \texttt{ChemicalReactionNetworkModels.jl}.


In situations where the number of molecules in the system is sufficiently large, 
the forwards evolution of the biochemical reaction network can be accurately approximated by the \emph{chemical Langevin equation}~\citep{Higham2008,Gillespie2000,Wilkinson2009}. The chemical Langevin equation is an It\={o} SDE of the form
\begin{equation}
\text{d}\bvec{X}_t = \sum_{j=1}^M \boldsymbol{\nu}_ja_j(\bvec{X}_t)\text{d}t + \sum_{j=1}^M \boldsymbol{\nu}_j\sqrt{a_j(\bvec{X}_t)}\text{d}W_t^{(j)}, \label{eq:CLE}
\end{equation}
where $\bvec{X}_t$ takes values in $\mathbb{R}^N$ and $W_t^{(1)},W_t^{(2)},\ldots,W_t^{(M)}$ are independent scalar Wiener processes. For a fixed initial condition, $\bvec{X}_0$, the solution to \eqref{eq:CLE}, $\{\bvec{X}_t\}_{0\leq t}$, can be approximately simulated using numerical methods. In this work, we apply the Euler-Maruyama scheme~\citep{Kloeden1999,Maruyama1955} which approximates a realisation at $\bvec{X}_{t+\Delta t}$ given $\bvec{X}_t$ according to
\begin{equation*}
\bvec{X}_{t+\Delta t} = \bvec{X}_t + \sum_{j=1}^M\boldsymbol{\nu}_ja_j(\bvec{X}_t) \Delta t + \sum_{j=1}^M \boldsymbol{\nu}_j\sqrt{a_j(\bvec{X}_t)\Delta t} \xi^{(j)},
\end{equation*} 
where $\xi^{(1)},\xi^{(2)},\ldots,\xi^{(M)}$ are independent, identically distributed (i.i.d.) standard normal random variables. It can be shown that the Euler-Maruyama scheme converges with rate $\mathcal{O}(\sqrt{\Delta t})$ to the true path-wise solution~\citep{Kloeden1999}. While higher-order schemes are possible, tighter restrictions on the SDE form are required. Therefore we restrict ourselves to Euler-Maruyama in this work. For an accessible introduction to numerical methods for SDEs, see \citet{Higham2001}, and for a detailed monologue that includes rigorous analysis of convergence rates, see \cite{Kloeden1999}. Example implementations of the Euler-Maruyama scheme for the chemical Langevin equation are provided in \texttt{EulerMaruyama.jl} and \texttt{ChemicalLangevin.jl}.

\subsection{Markov chain Monte Carlo for Bayesian inference}
\label{sec:intro_mcmc}

In practice, the application of mathematical models to the study of real biochemical networks requires model calibration and parameter inference using experimental data. The data are typically chemical concentrations derived from optical microscopy and fluorescent reporters such as green fluorescent proteins~\citep{Finkenstadt2008,Sahl2017,Wilkinson2011}. Let $\paramvec \in \paramspace$ be the vector of unknown model parameters, such as kinetic rate parameters or initial conditions. The task is to quantify the uncertainty in model parameters after taking the experimental data, $\dat$, into account. Given a model parameterised by $\paramvec$ and experimental data, $\dat$, uncertainty of the unknown model parameters can be quantified using the Bayesian \emph{posterior probability density},
\begin{equation}
\CondPDF{\paramvec}{\dat} = \frac{\like{\paramvec}{\dat}\PDF{\paramvec}}{\PDF{\dat}}, \label{eq:bayes}
\end{equation}
where: $\PDF{\paramvec}$ is the \emph{prior probability density} that encodes parameter assumptions; $\like{\paramvec}{\dat}$ is the likelihood function that determines the probability of the data under the assumed model for fixed $\paramvec$; and $\PDF{\dat}$ is the \emph{evidence} that provides a total probability for the data under the assumed model over all possible parameter values.

Parameter uncertainty quantification often involves computing expectations of functionals with respect to the posterior distribution (\eqref{eq:bayes}),
\begin{equation*}
\E{f(\paramvec)} = \int_{\paramspace} f(\paramvec)\CondPDF{\paramvec}{\dat}\, \text{d}\paramvec,
\end{equation*}
which may be estimated using Monte Carlo integration,
\begin{equation*}
\E{f(\paramvec)} \approx \hat{f}(\paramvec) = \frac{1}{\mathcal{M}}\sum_{i=1}^{\mathcal{M}}f(\paramvec^{(i)}), 
\end{equation*}
where $\paramvec^{(1)},\paramvec^{(2)}, \ldots, \paramvec^{(\mathcal{M})}$ are i.i.d.~samples from the posterior distribution, $\CondPDF{\paramvec}{\dat}$. In particular, the $j$th marginal posterior probability density, $\CondPDF{\param_j}{\dat}$ with $\param_j$ the $j$th dimension of $\paramvec$, may be estimated using a smoothed kernel density estimate,
\begin{equation*}
\CondPDF{\param_j}{\dat} \approx \frac{1}{\mathcal{M}h} \sum_{i=1}^\mathcal{M} K\left(\frac{\param_j - \param_j^{(i)}}{h}\right),
\end{equation*}
where  $\param_j^{(1)},\param_j^{(2)}, \ldots, \param_j^{(\mathcal{M})}$ are the $j$th dimensions of i.i.d. posterior samples, $h$ is a user prescribed smoothing parameter, and the kernel $K(x)$ is chosen such that $\int_{-\infty}^{\infty}K(x)\,\text{d}x = 1$. Typically, $K(x)$ is a standard Gaussian, and $h$ is chosen using Silverman's rule~\citep{Silverman1986}.
In most cases, direct i.i.d.~sampling from the posterior distribution is not possible since it is often not from a standard distribution family.

MCMC methods are based on the idea of simulating a discrete time Markov chain, $\{\paramvec_m\}_{0\leq m}$, in parameter space, $\paramspace$, for which the posterior of interest is its stationary distribution~\citep{Green2015,Roberts2004}. A popular MCMC algorithm is the Metropolis-Hastings method~\citep{Hastings1970,Metroplis1953} (Algorithm~\ref{alg:mcmcMH}, an example implementation is provided in \texttt{MetropolisHastings.jl}), in which transitions from state $\paramvec_{m}$ to a proposed new state $\paramvec^*$ occur with probability proportional to the relative posterior density between the two locations. 
\begin{algorithm}
	\caption{The Metropolis-Hastings method for MCMC}
	\begin{algorithmic}[1]
		\State{Given initial condition $\paramvec_0$ such that $\CondPDF{\paramvec_0}{\dat} > 0$.}
		\For{$m = 1, \ldots, \mathcal{M}$}
		\State Sample transition kernel, $\paramvec^* \sim \Kernel{\paramvec}{\paramvec_{m-1}}$.
		\State Calculate acceptance probability
		\begin{equation*}
		\alpha\left(\paramvec^*,\paramvec_{m-1}\right) = \min \left(1, \dfrac{ \Kernel{\paramvec_{m-1}}{\paramvec^*}\CondPDF{\paramvec^*}{\dat}}{\Kernel{\paramvec^{*}}{\paramvec_{m-1}}\CondPDF{\paramvec_{m-1}}{\dat}}\right).
		\end{equation*}
		\State Set $\paramvec_{m} \leftarrow \paramvec^{*}$ with probability $\alpha\left(\paramvec^*,\paramvec_{m-1}\right)$, otherwise, set $\paramvec_{m} \leftarrow \paramvec_{m-1}$.
		
		\EndFor
	\end{algorithmic}
	\label{alg:mcmcMH}
\end{algorithm}
The proposals are determined though sampling a proposal kernel distribution that is conditional on $\paramvec_{m}$ with density $\Kernel{\paramvec^*}{\paramvec_m}$. Under some regularity conditions on the proposal density, the resulting Markov chain will converge to the target posterior as its stationary distribution~\citep{Mengersen1996}.   
Therefore, computing expectations can be performed with Monte Carlo integration using a sufficiently large \emph{dependent} sequence from the Metropolis-Hastings Markov chain. It is important to note that this is an asymptotic result, and determining when such a sequence is sufficiently large for practical purposes is an active area of research~\citep{Cowles1996,Gelman1992,Gelman2014,Vehtari2019}.  It is also important to note that the efficiency of MCMC based on Metropolis-Hastings is heavily dependent on the proposal density used~\citep{Metroplis1953}. Adaptive schemes may be applied~\citep{Roberts2009}, however, care must be taken when applying these schemes as the stationary distribution may be altered. In many practical applications, the proposal density and number of iterations is selected heuristically~\citep{Hines2014}.

Alternative MCMC algorithms include Gibbs sampling~\citep{Geman1984}, Hamiltonian Monte Carlo~\citep{Duane1987}, and Zig-Zag sampling~\citep{Bierkens2019}. In this work, however, we base all discussion and examples on the Metropolis-Hastings method as it is the most natural to extend to challenging inference problems in systems biology~\citep{Golightly2011,Marjoram2003}.  

\subsection{A tractable example: the production-degradation model}
\label{sec:intro_example}
We demonstrate the application of MCMC to perform exact Bayesian inference using a biochemical reaction network for which an analytic solution to the likelihood can be obtained. This enables us to highlight important MCMC algorithm design considerations before introducing the additional complexity that arises when the likelihood is intractable.

\begin{figure}
	\centering
	\includegraphics[width=1\linewidth]{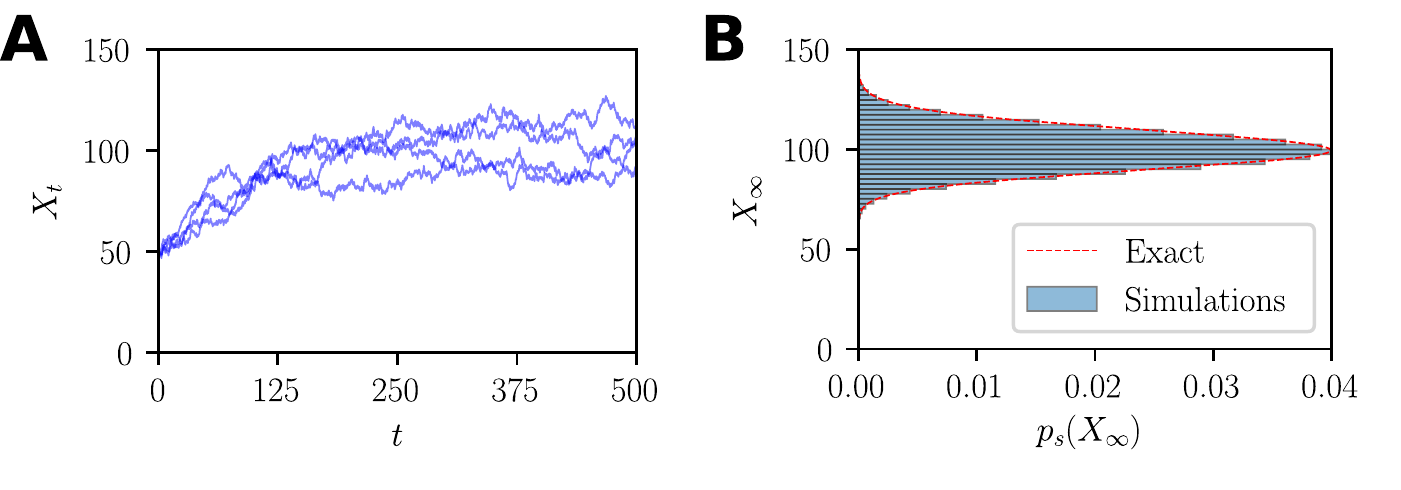}
	\caption{(A) Four example realisations of the chemical Langevin SDE for the production-degradation model. (B) The analytic stationary distribution compared with an approximation using simulations. Simulations are performed using the Euler-Maruyama scheme with $\Delta t = 0.1$ and $X_0 = 50$. Kinetic rate parameters are $k_1 = 1.0$ and $k_2 = 0.01$.}
	\label{fig:fig1}
\end{figure}

Consider a biochemical system consisting a single chemical species, $X$, involving only production and degradation reactions of the form 
\begin{equation}
\underbrace{\emptyset \overset{k_1}{\rightarrow} X}_{\substack{\text{external production}\\ \text{ of $X$ molecules}}} \quad \text{and}\quad  \underbrace{X \overset{k_2}{\rightarrow} \emptyset}_{\substack{\text{degradation}\\ \text{of $X$ molecules}}}. \label{eq:proddeg}
\end{equation} 
Here, $k_1 > 0$ and $k_2 > 0$ are the kinetic parameters for production and degradation, respectively. The propensity functions are given by
\begin{equation*}
a_1(X_t) = k_1\quad \text{and}\quad a_2(X_t) = k_2X_t,
\end{equation*}
with respective stoichiometries $\nu_1 = 1$ and $\nu_2 = -1$. The chemical Langevin equation for this production-degradation model (\eqref{eq:proddeg}) is 
\begin{equation}
\text{d}X_t = (k_1 - k_2 X_t)\text{d}t + \sqrt{k_1 + k_2X_t}\text{d}W_t, \label{eq:CLEpd}
\end{equation}
where $W_t$ is a Wiener process.  Approximate realisations of the solution process can be generated using the Euler-Maruyama discretisation, as demonstrated in Figure~\ref{fig:fig1}(A) (see example \texttt{DemoProdDeg.jl}). Throughout this work we take time, $t$, and rate parameters to be dimensionless. However, all results can be re-dimensionalised as appropriate.

Assume that the degradation rate is known, $k_2 = 0.01$. The inference task is to quantify the uncertainty in the production kinetic rate, $k_1$, using experimental data $\dat = \left[ Y_\text{obs}^{(1)},Y_\text{obs}^{(2)},\ldots, Y_\text{obs}^{(n)}\right]$, where $Y_\text{obs}^{(1)},Y_\text{obs}^{(2)},\ldots, Y_\text{obs}^{(n)}$ are $n$ independent observations of a hypothetical real biochemical production-degradation process that has reached its equilibrium distribution (Appendix~\ref{sec:obs_data}). For simplicity, we also assume these observations are not subject to any observation error, that is, our data is assumed to be exact realisations of the stationary process for the production-degradation model (\eqref{eq:proddeg}) under the chemical Langevin equation representation (\eqref{eq:CLEpd}). 

For inference, we require the Bayesian posterior probability density, 
\begin{equation}
\CondPDF{k_1}{\mathcal{D}} \propto \like{k_1}{\dat}\PDF{k_1}, \label{eq:postpd}
\end{equation}
where the prior is $\PDF{k_1}$ and the likelihood is
\begin{equation}
\like{k_1}{\dat} = \prod_{i=1}^n p_s\left(Y_{\text{obs}}^{(i)};k_1\right).\label{eq:likepd}
\end{equation}
We prescribe a uniform prior, $k_1 \sim \mathcal{U}(0,2)$, that contains the true parameter value of $k_1 = 1.0$. In~\eqref{eq:likepd},  $p_s\left(x; k_1\right)$ is the probability density function for the chemical Langevin equation (\eqref{eq:CLEpd}) solution process, $\{X_t\}_{0 \leq t}$, as $t \to \infty$, that is, the stationary process $X_\infty \sim p_s(x;k_1)$.
For this particular example, it is possible to obtain an analytical expression for this stationary probability density function. The solution is obtained by formulating the Fokker-Planck equation for the It\={o} process in \eqref{eq:CLEpd} and solving for the steady state (Appendix~\ref{sec:app_stat_pd}) to yield 
\begin{equation}
p_s\left(x; k_1\right) = \frac{\exp\left({-2x + \left(\dfrac{4k_1}{k_2}-1\right)\ln\left(k_1+k_2x\right)}\right)}{\displaystyle\int_{0}^{\infty}\exp\left({-2y + \left(\frac{4k_1}{k_2}-1\right)\ln\left(k_1+k_2y\right)}\right)\, \text{d}y}.
\label{eq:pdstat}
\end{equation}
Given a value for $k_1$, then the denominator can be accurately calculated using quadrature. Figure~\ref{fig:fig1}(B) overlays this analytical solution against a histogram obtained from the time series of a single very long simulation with end time, $t = 1,000,000$.

Using \eqref{eq:pdstat}, we can now evaluate the likelihood function (\eqref{eq:likepd} point-wise, and hence the posterior density (\eqref{eq:postpd}) can be evaluated point-wise up to a normalising constant. Therefore, we can apply the Metropolis-Hastings method for which the acceptance probability is
\begin{equation}
\alpha\left(\param^*,\param\right) = \min\left(1,\frac{\Kernel{\param}{\param^*}p(\param^*)\prod_{i=1}^n p_s\left(Y_{\text{obs}}^{(i)};\param^*\right)}{\Kernel{\param^*}{\param}p(\param)\prod_{i=1}^n p_s\left(Y_{\text{obs}}^{(i)};\param\right)}\right), \label{eq:accpd}
\end{equation}
where $\param^* \sim \Kernel{\cdot}{\param}$ is the proposal mechanism. 

\begin{figure}
	\centering
	\includegraphics[width=0.9\linewidth]{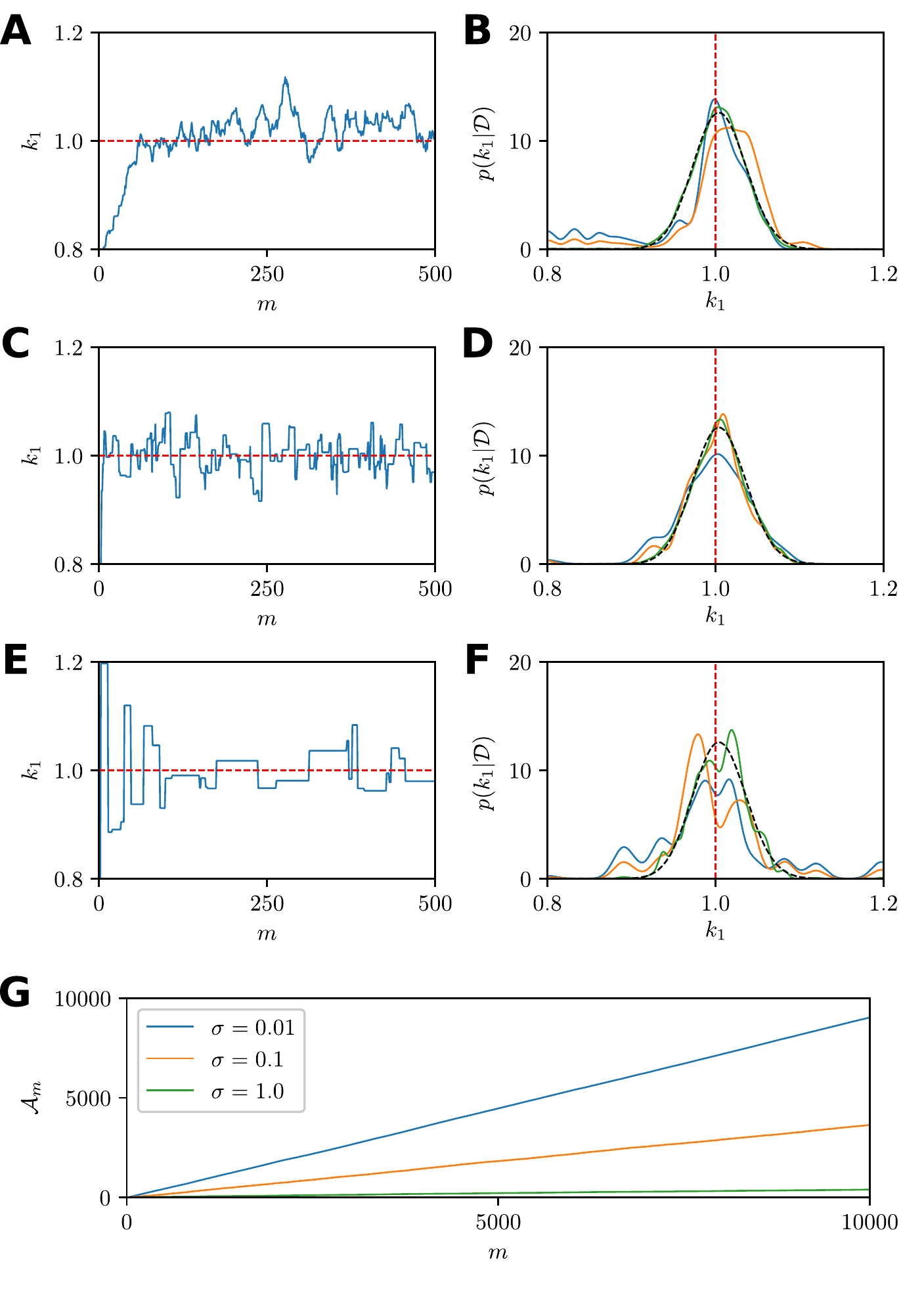}
	\caption{The choice of proposal kernel affects the convergence of the Markov chain. (A, C, and E) Trace plots for the first 500 iterations production kinetic parameter $k_1$, and (B, D, and F) smoothed kernel density estimates for the target Bayesian posteriors are shown for Gaussian proposal kernels with standard deviations of: (A-B) $\sigma = 0.01$; (C-D) $\sigma = 0.1$ and (E-F) $\sigma = 1.0$. Smoothed kernel density estimates for target Bayesian posterior are shown at 250 iterations (solid blue), 500 iterations (solid orange), and 10,000 iterations (solid green) alongside the exact posterior (dashed black). The true production rate of $k_1 = 1.0$ is indicated (dashed red). (G) Comparison of cumulative accepted proposal counts $\mathcal{A}_m$ for different proposal variances, $\sigma = 0.01$ (solid blue), $\sigma = 0.1$ (solid orange) and $\sigma = 1.0$ (solid green). }
	\label{fig:fig2}
\end{figure}

The choice of proposal kernel dramatically affects the rate of convergence of the Markov chain. For example, Figure~\ref{fig:fig2} demonstrates the Markov chain based on \eqref{eq:accpd} using a Gaussian proposal kernel,
\begin{equation}
\Kernel{\param^*}{\param} = \frac{1}{\sigma\sqrt{2\pi}}\exp\left(-\frac{(\param^* - \theta)^2}{2\sigma^2}\right),
\end{equation} 
for different choices of the standard deviation parameter $\sigma$ (see example \texttt{DemoMH.jl}). In all cases, the initial state of the chain is set in a region of very low posterior density, $\param_0 = 0.8$, to ensure we can compare transient and stationary behaviour of the Markov chain. 
For small standard deviation, $\sigma = 0.01$ (Figure~\ref{fig:fig2}(A)--(B)), the move acceptance rate is high~(see Figure~\ref{fig:fig2}(G)), however, only very small steps are ever taken (Figure~\ref{fig:fig2}(A)). These small steps lead to an over sampling of the low density region of initialisation before the chain drifts toward the high density region. This over sampling of the tail is still evident after $500$ iterations (Figure~\ref{fig:fig2})(B)), and almost $10,000$ iterations are required to compensate for this initial transient behaviour. We emphasize that here we refer to transient behaviour of the Metropolis-Hastings Markov chain, and this is not to be confused with any transient behaviour of the underlying model. In Figure~\ref{fig:fig2}(C)--(D), we show that the use of a larger standard deviation, $\sigma = 0.1$, results in the rejection of significantly more proposals (Figure~\ref{fig:fig2}(C) and Figure~\ref{fig:fig2}(G)), however, the larger steps result in rapid convergence to the true target density in almost $500$ iterations (Figure~\ref{fig:fig2}(D)). However, increasing the standard deviation further to $\sigma = 1.0$ (Figure~\ref{fig:fig2}(E)--(F)), results in proposals that overshoot the high density region frequently and most proposals are rejected~(Figure~\ref{fig:fig2}(G)). Consequently, the chain halts for many iterations and parameter space exploration is inefficient (Figure~\ref{fig:fig2}). Even after $10,000$ iterations the chain still has not reached stationarity. 

In Figure~\ref{fig:fig2}(B),(D), and (F), the exact posterior density for the production-degradation inference problem, $\CondPDF{k_1}{\dat}$, is overlaid to demonstrate that all chains are still converging to the same stationary distribution. Having this exact solution also enables us to demonstrate the impact that the choice of proposal kernel has on the efficacy of the Metropolis-Hastings method. In general, the selection of an optimal proposal kernel is an open problem, however, there are techniques that can be applicable in specific cases~\citep{Gelman1996,Roberts2009,Yang2013}.  
\FloatBarrier

%

\section{Likelihood-free MCMC}

Nearly all likelihood functions for stochastic biochemical systems of interest are intractable. This renders the standard Metropolis-Hastings method for MCMC sampling (Algorithm~\ref{alg:mcmcMH}) impossible to implement directly~\citep{Sisson2018,Warne2019,Wilkinson2011}. To deal with this problem, techniques for sampling Bayesian posterior distributions have been developed that avoid the point-wise evaluation of the likelihood. These so-called \emph{likelihood-free} methods fall into two main categories: approximate Bayesian computation; and pseudo-marginal methods. 

In this section, we provide a brief description of both approaches in the context of the Metropolis-Hastings method for MCMC sampling. We then demonstrate some important features of these methods in the context of the tractable production-degradation inference problem presented in Section~\ref{sec:intro_example}. 

\subsection{Approximate Bayesian computation}
ABC is a broad class of Bayesian sampling techniques that are applicable when the likelihood is intractable but simulated data can be generated efficiently for a given parameter vector $\paramvec$~\citep{Sisson2018,Sunnaker2013,Warne2019}. The fundamental idea is that parameter values that frequently lead to simulated data $\simdat$ that are \emph{similar} to the true observations $\dat$ will have higher posterior probability density. In effect, ABC samples from an approximate Bayesian posterior, 
\begin{equation}
\CondPDF{\paramvec}{\discrep{\dat}{\simdat} \leq \epsilon} \propto \CondProb{\discrep{\dat}{\simdat} \leq \epsilon}{\paramvec}\PDF{\paramvec}, \label{eq:abcpost}
\end{equation}
where the \emph{discrepancy metric}, $\discrep{\dat}{\simdat}$, quantifies how different the two datasets are, the \emph{acceptance threshold}, $\epsilon > 0$, specifies the difference that is considered close, and $\simdat \sim \simProc{\dat}{\paramvec}$ is the data simulation process.

In the context of MCMC sampling, \citet{Marjoram2003} developed a modified Metropolis-Hastings method using the acceptance probability
\begin{equation}
\alpha(\paramvec^*,\paramvec_m) = \begin{cases}
\min\left(1, \dfrac{\Kernel{\paramvec_m}{\paramvec^*}\PDF{\theta^*}}{\Kernel{\paramvec^*}{\paramvec_m}\PDF{\theta_m}}\right), & \text{ if } \discrep{\dat}{\simdat} \leq \epsilon, \\
0, & \text{ if } \discrep{\dat}{\simdat} > \epsilon. \label{eq:acceptABC}
\end{cases}
\end{equation}
\citet{Marjoram2003} also show that the stationary distribution of the resulting Markov chain is~\eqref{eq:abcpost}. Provided that the discrepancy metric, $\discrep{\dat}{\simdat}$, and acceptance threshold, $\epsilon$, are appropriately selected so that $\CondPDF{\paramvec}{\discrep{\dat}{\simdat} \leq \epsilon} \approx \CondPDF{\paramvec}{\dat}$, then we can use this Markov chain for inference in the same way that the chain from classical Metropolis-Hastings MCMC (Algorithm~\ref{alg:mcmcMH}) would be used. An example implementation is provided in \texttt{ABCMCMC.jl}.

The choice of $\discrep{\dat}{\simdat}$ and $\epsilon$ are critical to both the accuracy of approximate posterior, and the computational cost of the method. Ideally, we require $\discrep{\dat}{\simdat}$ such that we recover the true posterior density in the limit as $\epsilon \to 0$. Using a metric such as the Euclidean distance satisfies this property, however, when the data has high dimensionality it is completely infeasible for small $\epsilon$ to accept any parameter proposals. Conversely, metrics based on summary statistics of the data can be used to reduce the data dimensionality so that a smaller $\epsilon$ can be used, however, this may not lead to the true posterior as $\epsilon \to 0$. In general, one requires the summary statistics to be sufficient statistics~\citep{Fearnhead2012} and $\epsilon$ to be of similar order to the observation error~\citep{Toni2008,Wilkinson2013} to obtain accurate posteriors for the purposes of inference.

\subsection{Pseudo-marginal methods}

Pseudo-marginal methods~\citep{Andrieu2009} are an alternative approach to likelihood-free inference with some desirable properties compared with ABC. The pseudo-marginal approach can be used when one has an unbiased Monte Carlo estimator, $\mclike{\theta}{\dat}$, for the point-wise evaluation of the likelihood function $\like{\theta}{\dat}$. This estimator is used directly in place of the true likelihood for the purposes of MCMC. 

In the context of Metropolis-Hastings MCMC (Algorithm~\ref{alg:mcmcMH}), the acceptance probability for the pseudo-marginal approach is
\begin{equation}
\alpha(\paramvec^*,\paramvec_m) = \min\left(1, \dfrac{\Kernel{\paramvec_m}{\paramvec^*}\mclike{\paramvec^*}{\dat}\PDF{\theta^*}}{\Kernel{\paramvec^*}{\paramvec_m}\mclike{\paramvec_m}{\dat}\PDF{\theta_m}}\right). \label{eq:acceptPM}
\end{equation}
After initial inspection, one would expect the stationary distribution of the resulting Markov chain to be an approximation to the true posterior, just as with the ABC approach using~\eqref{eq:acceptABC}. Surprisingly, this is not the case; the stationary distribution of the Markov chain using~\eqref{eq:acceptPM} is, in fact, the exact posterior distribution (\eqref{eq:bayes}). As a result, pseudo-marginal methods have been referred to as \emph{exact approximations}~\citep{Golightly2011}. For a brief explanation for why the true posterior is recovered, see Appendix~\ref{sec:app_pm_exact}. For more detail we refer the reader to~\cite{Andrieu2009}, \cite{Beaumont2003}, and \cite{Golightly2011}. An example implementation is provided in \texttt{PseudoMarginalMetropolisHastings.jl}.

Unlike classical Metropolis Hastings, the acceptance probability, $\alpha(\paramvec^*,\paramvec_m)$ (\eqref{eq:acceptPM}), is still a random variable, given values for $\paramvec^*$ and $\paramvec_m$. This additional randomness reduces the rate at which the Markov chain approaches stationarity, but the additional noise can be controlled through reducing the variance of the likelihood estimator $\mclike{\paramvec}{\dat}$. However, reducing the variance necessarily requires higher computation costs since a larger number of Monte Carlo samples will be required for computing $\mclike{\paramvec}{\dat}$. Research has been undertaken to try to develop methods to choose the number of samples optimally. In particular, \citet{Doucet2015} perform a detailed analysis and find, under some restrictive assumptions, that the choosing the number of samples such that $\V{\log \mclike{\bar{\paramvec}}{\dat}} \approx 1.2$, where $\bar{\paramvec}$ is the posterior mean, is the optimal trade-off.  

\subsection{Comparison for an example with a tractable likelihood}
We now demonstrate the ABC and pseudo-marginal approaches to MCMC using the tractable production-degradation problem from Section~\ref{sec:intro_example}. Specifically, we demonstrate how the ABC acceptance threshold and the pseudo-marginal Monte Carlo estimator variance affect both the rate of convergence and the stationary distribution.

\begin{landscape}
	\begin{figure}
		\centering
		\includegraphics[width=0.9\linewidth]{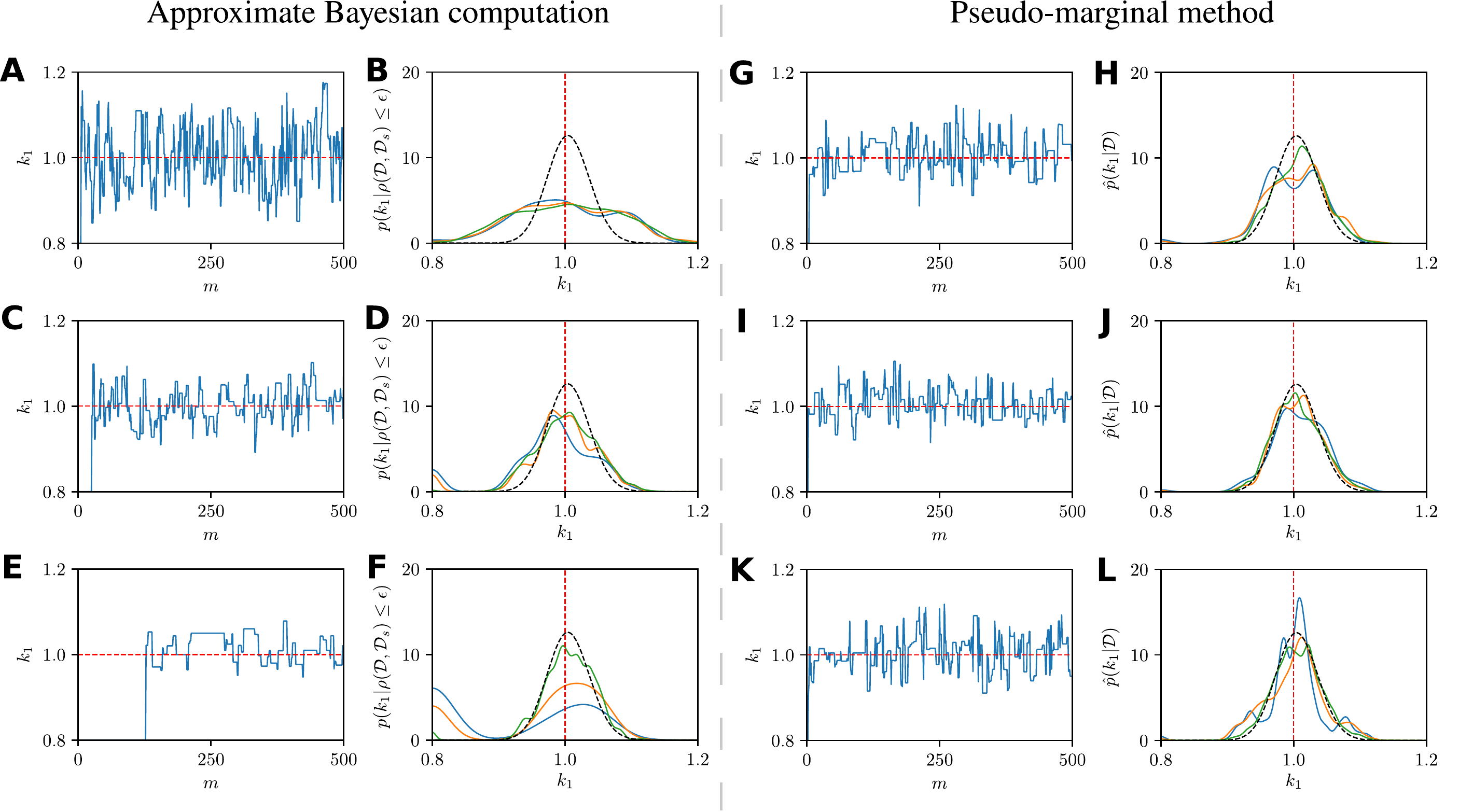}
		\caption{Comparison of likelihood-free MCMC algorithms, (A-F) ABC and (G-L) the pseudo-marginal approach, using the tractable production degradation example. ABC trace plots and smoothed kernel density estimates for the target approximate Bayesian posterior are shown for decreasing acceptance thresholds: (A-B) $\epsilon = 14$; (C-D) $\epsilon = 7$ and (E-F) $\epsilon = 3.5$. Pseudo-marginal trace plots and smoothed kernel density estimates for the target approximate Bayesian posterior are shown for increasing sample numbers for likelihood estimation: (G-H) $R = 25$; (I-J) $R = 50$ and (K-L) $R = 100$.  Smoothed kernel density estimates for target Bayesian posterior are shown at 250 iterations (solid blue), 500 iterations (solid orange), and 10,000 iterations (solid green) alongside the exact posterior (dashed black). The true production rate of $k_1 = 1.0$ is indicated (dashed red). In all cases, the proposal kernel is Gaussian with variance $\sigma^2 = 0.01$, and the chain is initialised with $\param_0 = 0.8$.}
		\label{fig:fig4}
	\end{figure}
\end{landscape}

For the ABC case, we can generate simulated data $\simdat = \left[X_T^{(1)},X_T^{(2)},\ldots, X_T^{(n)}\right]$ where\\ $X_T^{(1)},X_T^{(2)},\ldots,X_T^{(n)}$ are independent approximate realisations of the production degradation model (\eqref{eq:CLEpd}) using the Euler-Maruyama scheme over the interval $0 \leq t \leq T$ with $\Delta t = 1.0$, $T = 1000.0$, $k_2 = 0.01$, and $k_1$ is given by the state of the Markov chain $\param_m$. Given that the data has dimension $n = 10$ (Section~\ref{sec:intro_example},  Appendix~\ref{sec:obs_data}), we choose a discrepancy metric that reduces the data dimension for ease of demonstration, that is,  
\begin{equation*}
\discrep{\dat}{\simdat} = \left|\hat{\mu}(\dat)  - \hat{\mu}(\simdat)\right| + \left|\hat{\sigma}(\dat) - \hat{\sigma}(\simdat)\right|,
\end{equation*} 
where $\hat{\mu}(\dat)$ and $\hat{\sigma}(\dat)$ are the sample mean and standard deviation of the observations $\dat$, and  $\hat{\mu}(\simdat)$ and $\hat{\sigma}(\simdat)$ are the sample mean and standard deviation of the simulated data $\simdat$. Figure~\ref{fig:fig4}(A)--(F) demonstrates the behaviour of ABC MCMC using acceptance thresholds of $\epsilon = 14$ (Figure~\ref{fig:fig4}(A)--(B)), $\epsilon = 7$ (Figure~\ref{fig:fig4}(C)--(D)) and $\epsilon = 3.5$ (Figure~\ref{fig:fig4}(E)--(F)) (see example \texttt{DemoABCMCMC.jl}). The Markov chain trajectories shown in Figure~\ref{fig:fig4}(A), (C), (E), indicate that larger values of $\epsilon$ lead to more rapid convergence to stationarity. The converse is true for the accuracy of the stationary distribution as an approximation to the exact posterior, as demonstrated in Figure~\ref{fig:fig4}(B), (C), (F), with larger values of $\epsilon$ leading to a more diffuse, approximate posterior. This highlights a known shortcoming for ABC for the purposes of MCMC sampling; choosing a small $\epsilon$ for accuracy will tend to result in a Markov chain that repeatedly gets stuck in the same location~\citep{Sisson2007}.

For the pseudo-marginal approach we use a standard smoothed kernel density estimate for the likelihood, that is,
\begin{equation*}
\mclike{\paramvec}{\dat} = \frac{1}{(Rh)^n}\prod_{i=1}^n \sum_{j=1}^R K\left(\frac{Y_{\text{obs}}^{(i)} - X_T^{(j)}}{h}\right), 
\end{equation*}
where $h$ is the smoothing parameter chosen using Silverman's rule~\citep{Silverman1986}, $K(x)$ is a standard Gaussian smoothing kernel, and $X_T^{(1)},X_T^{(2)},\ldots,X_T^{(R)}$ are independent approximate realisations of the production degradation model (\eqref{eq:CLEpd}) using the Euler-Maruyama scheme with identical parameterisation as used for ABC. The variance of the likelihood estimator depends on the number of realisations used in the estimate, $R$. Figure~\ref{fig:fig4}(G)--(L) demonstrates the behaviour of the pseudo-marginal approach to MCMC using different realisation numbers of $R = 25$ (Figure~\ref{fig:fig4}(G)--(H)), $R = 50$ (Figure~\ref{fig:fig4}(I)--(J)) and $R = 100$ (Figure~\ref{fig:fig4}(K)--(L)) (see example \texttt{DemoPMMH.jl}). As expected, increasing $R$ has the effect of increasing convergence (although not significantly so). More importantly, regardless of the value $R$, the same stationary distribution is approached in the limit, that is, the exact posterior distribution.

This highlights a major advantage of pseudo-marginal methods, that is, the stationary distribution is independent of the number of realisations, $R$; furthermore the stationary distribution is the exact Bayesian posterior distribution. Even with $R = 1$ the method will eventually converge to the exact posterior distribution. This is in stark contrast to ABC methods where the stationary distribution depends on the discrepancy metric and acceptance threshold. Ultimately this means, that user choices only affect the computational performance of pseudo-marginal methods rather than both computational performance and inference accuracy with ABC. This is a clear advantage of the pseudo-marginal approach.


\section{Pseudo-marginal methods for biochemical systems}

The production-degradation example presented in Section~\ref{sec:intro_example} is useful for highlighting the essential concepts of standard MCMC sampling and likelihood free alternatives. However, this inference problem is very simple compared to practical problems, since real biochemical processes are generally not observed in their stationary state without observation error. Rather, real biochemical process data, as shown in Figure~\ref{fig:data}, is often characterised by noisy, time-course data, with few observations in time and only partially observed states~\citep{Finkenstadt2008,Golightly2011,Warne2019}. 

\subsection{The challenge for time-course data}
In the case of time-course data, the observations are samples at discrete points in time, $t_0, t_1, \ldots, t_n$, from a single realisation of a stochastic process, such as a gene regulatory network. The observation process, denoted by  $\{\bvec{Y}_t\}_{0 \leq t}$, often has the form,  $\bvec{Y}_t \sim \obsProc{\bvec{Y}_t}{\bvec{X}_t}$,
where $\{\bvec{X}_t\}_{0 \leq t}$ is the underlying stochastic process, prescribed by the chemical Langevin equation (\eqref{eq:CLE}), that governs the biochemical kinetics, and $\obsProc{\bvec{Y}_t}{\bvec{X}_t}$ is the observation process. The resulting discrete observations will be $\dat = [\bvec{Y}_{\text{obs}}^{(0)},\bvec{Y}_{\text{obs}}^{(1)},\ldots,\bvec{Y}_{\text{obs}}^{(n)}]$ with $\bvec{Y}_{\text{obs}}^{(i)} = \bvec{Y}_{t_i}$ for $i = 0,1,\ldots,n$. The likelihood for such observations is
\begin{equation}
\like{\paramvec}{\dat} = \PDF{\bvec{Y}_{\text{obs}}^{(0)}}\prod_{i=1}^n \CondPDF{\bvec{Y}_{\text{obs}}^{(i)}}{\bvec{Y}_{\text{obs}}^{(0)},\ldots,\bvec{Y}_{\text{obs}}^{(i-1)}}. \label{eq:likets}
\end{equation} 
Not only is this likelihood intractable, but a direct Monte Carlo likelihood estimator will be impractical for the pseudo-marginal approach. For example, the following is a direct Monte Carlo estimate for~\eqref{eq:likets}
\begin{equation}
\mclike{\paramvec}{\dat} = \frac{1}{R}\sum_{j=1}^R \prod_{i=1}^n \obsProc{\bvec{Y}_{\text{obs}}^{(i)}}{\bvec{X}_{t_i}^{(j)}}, \label{eq:est1}
\end{equation}
where $[\bvec{X}_{t}^{(1)},\bvec{X}_{t}^{(2)},\ldots,\bvec{X}_{t}^{(R)}]$ are $R$ independent realisations of the continuous sample path from the chemical Langevin equation (\eqref{eq:CLE}), that are subsequently observed at times $t_0, t_1, \ldots, t_n$. However, a prohibitively large number of sample paths, $R$, will be required to obtain an acceptable variance in the estimator in \eqref{eq:est1}. Consequently, more advanced approaches to pseudo-marginal are required. 

The following observation assists finding an alternative solution,
\begin{equation*}
\CondPDF{\bvec{Y}_{\text{obs}}^{(i)}}{\bvec{Y}_{\text{obs}}^{(0)},\ldots,\bvec{Y}_{\text{obs}}^{(i-1)}} = \int_{\mathbb{R}^N} \obsProc{\bvec{Y}_{\text{obs}}^{(i)}}{\bvec{X}_{t_i}}\CondPDF{\bvec{X}_{t_i}}{\bvec{Y}_{\text{obs}}^{(0)},\ldots,\bvec{Y}_{\text{obs}}^{(i-1)}}\, \text{d}\bvec{X}_{t_i}. 
\end{equation*}
That is, provided we are able to sample from $\CondPDF{\bvec{X}_{t_i}}{\bvec{Y}_{\text{obs}}^{(0)},\ldots,\bvec{Y}_{\text{obs}}^{(i-1)}}$ for all\\ $i = 1,2,\ldots, n$, then we can use the alternative Monte Carlo estimator
\begin{equation}
\mclike{\paramvec}{\dat} =  \prod_{i=1}^n \frac{1}{R}\sum_{j=1}^R \obsProc{\bvec{Y}_{\text{obs}}^{(i)}}{\bvec{X}_{t_i}^{(j)}}, \label{eq:est2}
\end{equation}
where $[\bvec{X}_{t_i}^{(1)},\bvec{X}_{t_i}^{(2)},\ldots,\bvec{X}_{t_i}^{(R)}]$ are $R$ samples from the distribution $\CondPDF{\bvec{X}_{t_i}}{\bvec{Y}_{\text{obs}}^{(0)},\ldots,\bvec{Y}_{\text{obs}}^{(i-1)}}$. This estimator will have lower variance because conditioning the samples $[\bvec{X}_{t_i}^{(1)},\bvec{X}_{t_i}^{(2)},\ldots,\bvec{X}_{t_i}^{(R)}]$ on all observations taken up to time $t_{i-1}$ automatically removes contributions by trajectories that do not match the observational history. The challenge is in the sampling of $\CondPDF{\bvec{X}_{t_i}}{\bvec{Y}_{\text{obs}}^{(0)},\ldots,\bvec{Y}_{\text{obs}}^{(i-1)}}$ and it motivates the use of, so called, \emph{particle filters}~\citep{Doucet2011}. We present the mathematical basis for this approach in the next section, along with practical examples that demonstrate how the method works in practice.

\subsection{Particle MCMC}

The bootstrap particle filter~\citep{Gordon1993,Doucet2011} is a technique based on sequential importance sampling~\citep{DelMoral2006}. This enables one to sample from the sequence of distributions $\CondPDF{\bvec{X}_{t_1}}{\bvec{Y}_{\text{obs}}^{0}},\CondPDF{\bvec{X}_{t_2}}{\bvec{Y}_{\text{obs}}^{0},\bvec{Y}_{\text{obs}}^{1}},$ $\ldots, \CondPDF{\bvec{X}_{t_n}}{\bvec{Y}_{\text{obs}}^{0},\ldots,\bvec{Y}_{\text{obs}}^{n-1}}$ and thereby evaluate the lower variance likelihood estimator~(\eqref{eq:est2}).

Suppose we have independent samples, called particles,
\begin{equation*}
\bvec{X}_{t_{i-1}}^{(1)},\bvec{X}_{t_{i-1}}^{(2)},\ldots,\bvec{X}_{t_{i-1}}^{(R)} \sim \CondPDF{\bvec{X}_{t_{i-1}}}{\bvec{Y}_{\text{obs}}^{0},\ldots,\bvec{Y}_{\text{obs}}^{i-1}}.
\end{equation*}
Then, using the Euler-Maruyama scheme (or similar), we can simulate each particle forward to time $t_i$. This results in a new set of independent particles
\begin{equation*}
\tilde{\bvec{X}}_{t_{i}}^{(1)},\tilde{\bvec{X}}_{t_{i}}^{(2)},\ldots,\tilde{\bvec{X}}_{t_{i}}^{(R)} \sim \CondPDF{\bvec{X}_{t_{i}}}{\bvec{Y}_{\text{obs}}^{0},\ldots,\bvec{Y}_{\text{obs}}^{i-1}}.
\end{equation*}
From these particles, we can evaluate the Monte Carlo estimate for the marginal likelihood at time $t_i$,
\begin{equation}
\hatCondPDF{\bvec{Y}_{\text{obs}}^{i}}{\bvec{Y}_{\text{obs}}^{0},\ldots,\bvec{Y}_{\text{obs}}^{i-1}} = \dfrac{1}{R}\sum_{k=1}^R\obsProc{\bvec{Y}_{\text{obs}}^{i}}{\tilde{\bvec{X}}_{t_i}^{(k)}}. \label{eq:marginallike}
\end{equation}

Provided we can then generate a new set of independent particles, 
\begin{equation*}
\bvec{X}_{t_{i}}^{(1)},\bvec{X}_{t_{i}}^{(2)},\ldots,\bvec{X}_{t_{i}}^{(R)} \sim \CondPDF{\bvec{X}_{t_{i}}}{\bvec{Y}_{\text{obs}}^{0},\ldots,\bvec{Y}_{\text{obs}}^{i}},
\end{equation*}
we can compute \eqref{eq:marginallike} for all $i = 1, 2, \ldots, n$, and hence compute the likelihood estimator given in \eqref{eq:est2}. Progress can be made by noting that, through application of Bayes' Theorem,
\begin{align*}
\CondPDF{\bvec{X}_{t_{i}}}{\bvec{Y}_{\text{obs}}^{0},\ldots,\bvec{Y}_{\text{obs}}^{i}} = \frac{\CondPDF{\bvec{Y}_{\text{obs}}^{i}}{\bvec{X}_{t_{i}}}\CondPDF{\bvec{X}_{t_{i}}}{\bvec{Y}_{\text{obs}}^{0},\ldots,\bvec{Y}_{\text{obs}}^{i-1}}}{\CondPDF{\bvec{Y}_{\text{obs}}^{i}}{\bvec{Y}_{\text{obs}}^{0},\ldots,\bvec{Y}_{\text{obs}}^{i-1}}}.
\end{align*}
Therefore, we can approximate the set of particles $\bvec{X}_{t_{i}}^{(1)},\bvec{X}_{t_{i}}^{(2)},\ldots,\bvec{X}_{t_{i}}^{(R)}$ by resampling the particles $\tilde{\bvec{X}}_{t_{i}}^{(1)},\tilde{\bvec{X}}_{t_{i}}^{(2)},\ldots,\tilde{\bvec{X}}_{t_{i}}^{(R)}$ with replacement using probabilities,
\begin{equation*}
\Prob{\bvec{X}_{t_{i}} = \tilde{\bvec{X}}_{t_{i}}^{(k)}} = \frac{\obsProc{\bvec{Y}_{\text{obs}}^{i}}{\tilde{\bvec{X}}_{t_{i}}^{(k)}}}{R\hatCondPDF{\bvec{Y}_{\text{obs}}^{i}}{\bvec{Y}_{\text{obs}}^{0},\ldots,\bvec{Y}_{\text{obs}}^{i-1}}} = \frac{\obsProc{\bvec{Y}_{\text{obs}}^{i}}{\tilde{\bvec{X}}_{t_{i}}^{(k)}}}{\sum_{j=1}^R\obsProc{\bvec{Y}_{\text{obs}}^{i}}{\tilde{\bvec{X}}_{t_{i}}^{(j)}}}.
\end{equation*} 
The result is a set of equally weighed particles approximately distributed according to\\ $\CondPDF{\bvec{X}_{t_{i}}}{\bvec{Y}_{\text{obs}}^{0},\ldots,\bvec{Y}_{\text{obs}}^{i}}$. This leads to the \emph{bootstrap particle filter} (Algorithm~\ref{alg:BPF})~\citep{Gordon1993}. An example implementation is provided in \texttt{BootstrapParticleFilter.jl}.
\begin{algorithm}
	\caption{The bootstrap particle filter for likelihood estimation}
	\begin{algorithmic}[1]
		\State{Initialise $i = 0$ and $\left\{\bvec{X}_{t_0}^{(k)}\right\}_{k=1}^R$ where $\bvec{X}_{t_0}^{(k)} \sim \CondPDF{\bvec{X}_{t_0}}{\bvec{Y}_{\text{obs}}^{0}}$ for $k = 1,2, \ldots, R$.}
		\For{$i = 1, \ldots, n$}
		\For{$k = 1,\ldots, R$}
		\State{Simulate particle forward, $\bvec{X}_{t_i}^{(k)} \sim \simProc{\bvec{X}_{t_i}^{(k)}}{\bvec{X}_{t_{i-1}}^{(k)}}$.}
		\State{Compute weight, $W_{i}^k \leftarrow \obsProc{\bvec{Y}_{\text{obs}}^{i}}{\bvec{X}_{t_i}^{(k)}}$.}
		\EndFor
		\State{Compute marginal likelihood estimate, $\hatCondPDF{\bvec{Y}_{\text{obs}}^{i}}{\bvec{Y}_{\text{obs}}^{0},\ldots,\bvec{Y}_{\text{obs}}^{i-1}} \leftarrow \dfrac{1}{R}\sum_{k=1}^R W_{i}^k$.}
		\State{Resample particles, $\left\{\bvec{X}_{t_i}^{(k)}\right\}_{k=1}^R$, with replacement using probabilities $W_i^k/\left[\sum_{j=1}^R W_i^j\right]$ for $k = 1,2, \ldots, R$.} 
		\EndFor
		\State{Compute likelihood estimate, $\mclike{\paramvec}{\dat} \leftarrow \prod_{i=1}^n \hatCondPDF{\bvec{Y}_{\text{obs}}^{i}}{\bvec{Y}_{\text{obs}}^{0},\ldots,\bvec{Y}_{\text{obs}}^{i-1}}$.}
	\end{algorithmic}
	\label{alg:BPF}
\end{algorithm}

\begin{figure}[h!]
	\centering
	\includegraphics[width=1\linewidth]{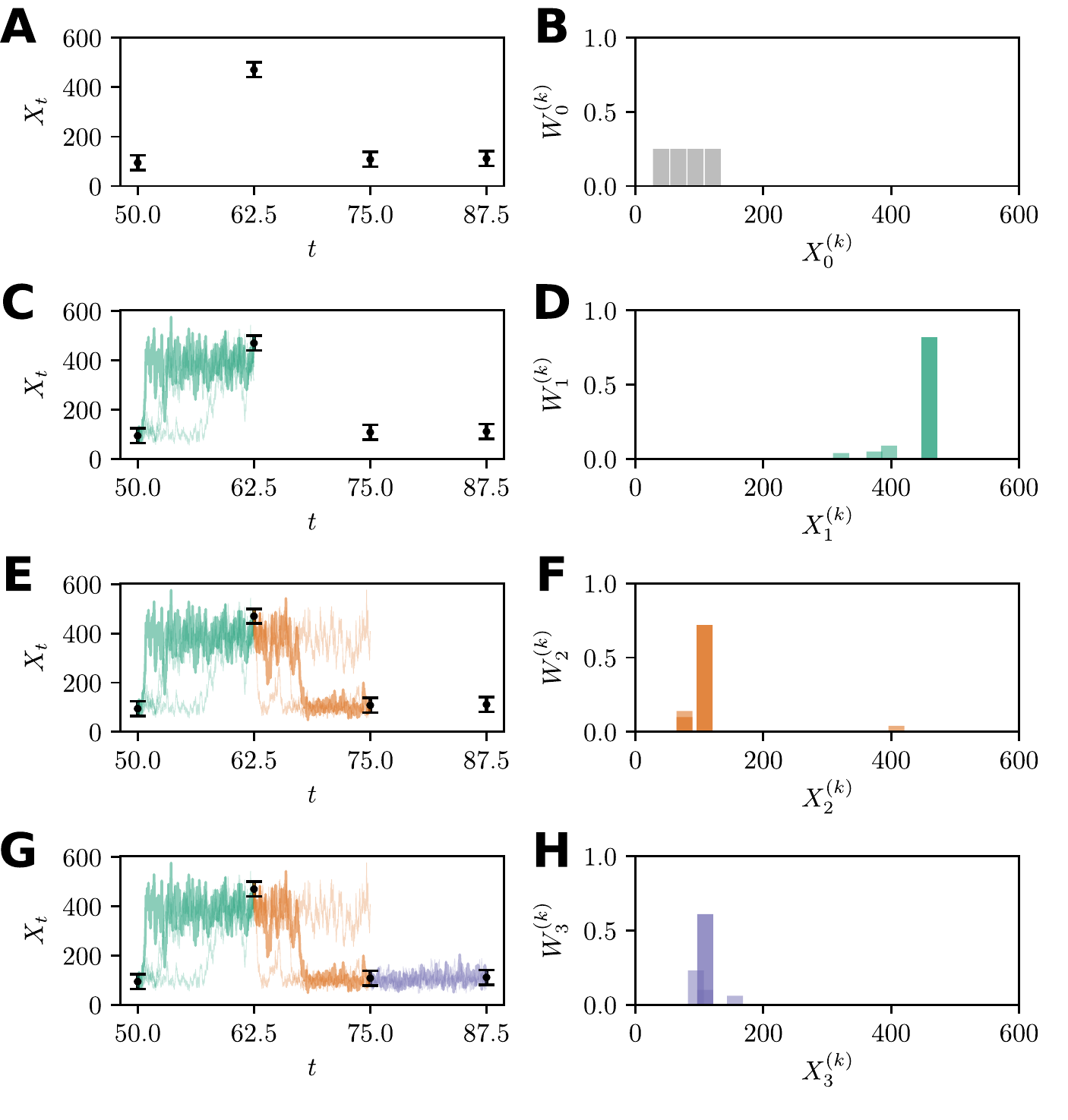}
	\caption{Demonstration of the bootstrap particle filter using $R =4$ particles for demonstration purposes. (A) Observations with error bars indication three standard deviations of the observation noise distribution. (B) Particles are initially weighted equally. (C)--(H) Three iterations of the boostrap particle filter. (C),(E), and (G) Particle forward trajectories with weights indicated by opacity. (D),(F), and (H)  Particle weight distributions computed for resampling.}
	\label{fig:fig5}
\end{figure}

Figure~\ref{fig:fig5} provides a visual demonstration of this process using a small number of particles, $R = 4$, for ease of visualisation. Figure~\ref{fig:fig5}(A) shows time-course data with error bars indicating the magnitude of the observation error. This data, with very low time resolution, is typical of many experimental studies, such as the data in Figure~\ref{fig:data}(C). Initially, all four particles are set to the initial data point with equal weighting (Figure~\ref{fig:fig5}(B)). The particles are then evolved forwards to the next observation time (Figure~\ref{fig:fig5}(C)), and the weighting is calculated as the probability density that the observation occurred based on each of these particles (Figure~\ref{fig:fig5}(D)). Note that only one of the four particles contribute significantly to the likelihood after this first forwards step, thus simulating the other three particles forwards any further would be a waste of computational effort. As such, we generate four independent continuations of the one highly weighted particle (Figure~\ref{fig:fig5}(D)) to evolve toward the next observation time (Figure~\ref{fig:fig5}(E)). The same process is repeated using the second generation of weights (Figure~\ref{fig:fig5}(F)), in order to perform the last step (Figure~\ref{fig:fig5}(G)-(H)).

The visualisation in Figure~\ref{fig:fig5} highlights the effect of conditioning on past observations. That is, model simulations that lead to a small likelihood in one of the observations are discarded early and new particles are generated by continuing high likelihood simulations. As a result all samples are focused on the higher density region of the likelihood and the variance of the estimator is reduced.

The application of particle filters for likelihood estimators within MCMC schemes is called \emph{particle MCMC}~\citep{Wilkinson2012}.  For the purposes for this article, we directly apply the pseudo-marginal method using Algorithm~\ref{alg:BPF} to evaluate the acceptance probability (\eqref{eq:acceptPM}) within the Metropolis-Hastings MCMC scheme (Algorithm~\ref{alg:mcmcMH}). This turns out to be a special case of the \emph{Particle marginal Metropolis-Hastings sampler} that may be used effectively to sample the joint probability density $\CondPDF{\paramvec,\bvec{X}_{t_0},\bvec{X}_{t_1},\ldots,\bvec{X}_{t_n}}{\bvec{Y}_\text{obs}^{0},\bvec{Y}_\text{obs}^{1},\ldots,\bvec{Y}_\text{obs}^{n}}$, as demonstrated within a very general framework introduced by \citep{Andrieu2010}.

\FloatBarrier
\subsection{Practical considerations}
There are a number of factors that may affect the performance of particle MCMC sampling in practice. Firstly, an important issue to discuss for sequential importance resampling, such as the bootstrap particle filter, is the problem of particle degeneracy. That is, as the number of iterations increases, the number of particles with non-zero weights decreases. As a result, the accuracy of approximation to $\CondPDF{\bvec{X}_{t_{i}}}{\bvec{Y}_{\text{obs}}^{0},\ldots,\bvec{Y}_{\text{obs}}^{i}}$ degrades as this dependency on a very small particle count introduces bias. While, the resampling step reduces the impact of degeneracy, in general a larger number of observations, $n$, will necessitate a large number of particles, $R$, for the likelihood estimator~\citep{Doucet2011}. The problem of degeneracy becomes even more problematic when the observation error is very small~\citep{Golightly2008,Golightly2011}, and may require more advanced resampling methods, such as systematic and stratified resampling~\citep{Kitagawa1996,Carpenter1999}. In this work, we apply a direct multinomial resampling scheme~\citep{Doucet2011}. 

Just as with the more general pseudo-marginal approach, there is a trade-off between the convergence rate of the Markov chain and the computational cost of each likelihood estimate. While~ 
\cite{Doucet2015} provide guidelines for optimally choosing $R$, these guides may not be feasible to implement since the behaviour of the likelihood estimator about the posterior mean is rarely known (especially since one is often using MCMC in order to compute this quantity).

The performance of particle MCMC methods also depends on the choice of proposal kernel, just as with classical Metropolis-Hastings. When the full inference problem is considered, there are a number of novel proposal schemes~\citep{Andrieu2010,Pooley2015}. A number of asymmetric proposal kernels, such as \emph{preconditioned Crank-Nicholson Langevin proposals}~\citep{Cotter2013}, can also be very effective in high dimensional parameter spaces. However, in general, one needs to perform experimentation to elucidate an effective combination of proposal kernel and particle numbers that will converge in an acceptable timeframe. 

The question of assessing convergence can be challenging. Typically, the auto-correlation functions (ACF) for each parameter are computed and the potential scale reduction is computed~\citep{Geyer1992}. However, these diagnostics for convergence can be very misleading, especially if the posterior is multimodal. To deal with this, it is common to use multiple chains and assess the within-chain and between-chain variances~\cite{Gelman1996,Gelman2014}. In this work, we follow the recent recommendations of~\citep{Vehtari2019}.

\section{Examples with intractable likelihoods}

As a practical demonstration of the use of particle MCMC, three examples are provided where expressions for the likelihood function are not available.

\subsection{Example 1: Michaelis-Menten enzyme kinetics}
The first example is based on the stochastic variant of the classical model for enzyme kinetics~\citep{Michaelis1913,Rao2003}.

\subsubsection{Model definition}
\begin{figure}[h]
	\centering
	\includegraphics[width=1\linewidth]{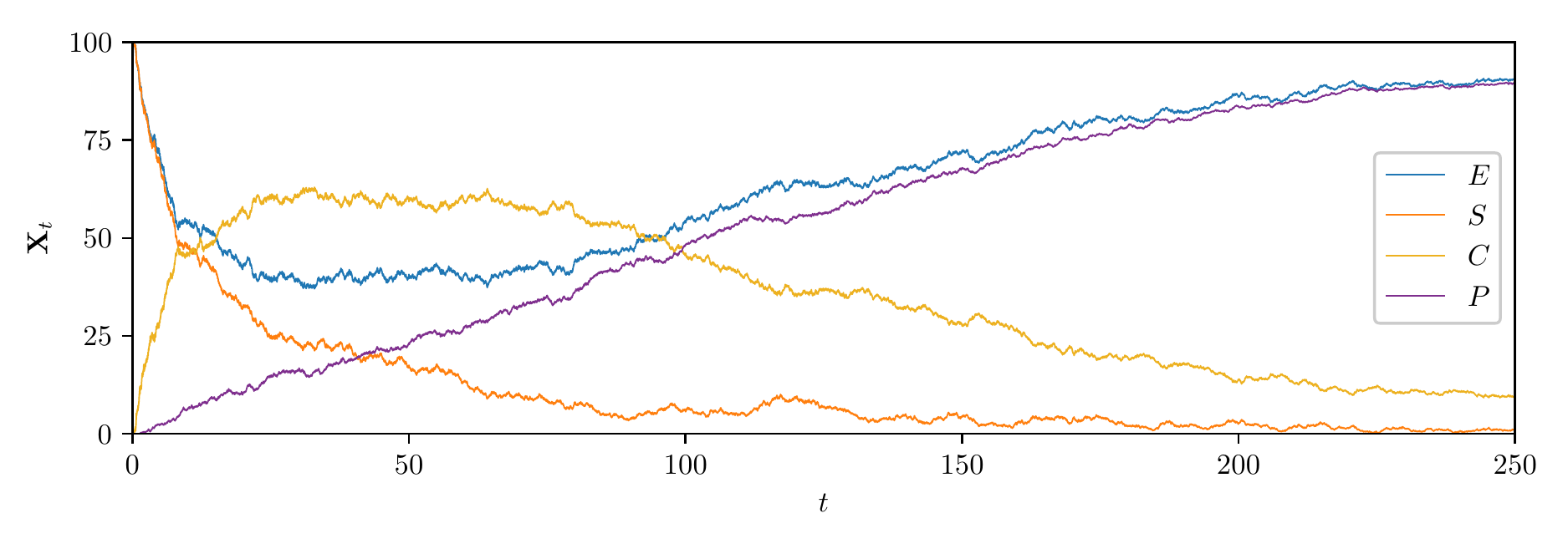}
	\caption{Example realisation of the Michaelis-Menten model demonstrating enzyme kinetics. The simulation is produced using the Euler-Maruyama scheme with $\Delta t = 1\times10^{-3}$, initial condition $\bvec{X}_0 = [100,100,0,0]$, and parameters $k_1 = 1\times10^{-3}$, $k_2 = 5\times10^{-3}$, and $k_3 = 1\times10^{-2}$.}
	\label{fig:fig6}
\end{figure}

The classical model of \cite{Michaelis1913} for enzyme kinetics describes the conversion of a chemical substrate $S$ into a product $P$ through the binding of an enzyme $E$. An enzyme molecule, $E$, binds to a substrate molecule, $S$, to form a complex, $C$, to convert $S$ to $P$. The stochastic process is describe through three reactions~\citep{Rao2003}, 
\begin{equation}
\underbrace{E + S \overset{k_1}{\rightarrow} C}_{\substack{\text{enzyme and substrate molecules}\\ \text{combine to form a complex}}},\quad  \underbrace{C \overset{k_2}{\rightarrow} E + S}_{\substack{\text{decay of complex}}}, \quad\text{and}\quad \underbrace{C \overset{k_3}{\rightarrow} E + P}_{\substack{\text{catalytic conversion}\\ \text{of substrate to product}}}, \label{eq:michment}
\end{equation}
with propensities, $a_1(\bvec{X}_t) = k_1 E_t S_t$,  $a_2(\bvec{X}_t) = k_2 C_t$, and $a_3(\bvec{X}_t) = k_3 C_t$,
where\\ $\bvec{X}_t = [E_t,S_t,C_t,P_t]^\text{T}$, and stoichiometries
\begin{equation*}
\boldsymbol{\nu}_1 = \begin{bmatrix}
-1 \\
-1 \\
1 \\
0
\end{bmatrix},\quad
\boldsymbol{\nu}_2 = \begin{bmatrix}
1 \\
1 \\
-1 \\
0
\end{bmatrix}\quad \text{and} \quad
\boldsymbol{\nu}_3 = \begin{bmatrix}
1 \\
0 \\
-1 \\
1
\end{bmatrix}.
\end{equation*}
Application of the chemical Langevin approximation (\eqref{eq:CLE}) to the Michaelis-Menten model (\eqref{eq:michment}) leads to a coupled system of It\={o} SDEs
\begin{equation}
\label{eq:CLEmichment}
\begin{split}
\text{d}E_t &= [-k_1E_t S_t + (k_2+k_3)C_t]\text{d}t - \sqrt{k_1E_tS_t}\text{d}W_t^{(1)} + \sqrt{k_2C_t}\text{d}W_t^{(2)} + \sqrt{k_3C_t}\text{d}W_t^{(3)},\\
\text{d}S_t &= (-k_1E_t S_2 + k_2 C_t)\text{d}t - \sqrt{k_1E_tS_t}\text{d}W_t^{(1)} + \sqrt{k_2C_t}\text{d}W_t^{(2)}, \\
\text{d}C_t &= [k_1E_t S_t - (k_2+k_3)C_t]\text{d}t + \sqrt{k_1E_tS_t}\text{d}W_t^{(1)} - \sqrt{k_2C_t}\text{d}W_t^{(2)} - \sqrt{k_3C_t}\text{d}W_t^{(3)},\\
\text{d}P_t &= k_3C_t\text{d}t +\sqrt{k_3C_t}\text{d}W_t^{(3)},
\end{split}
\end{equation}
where $W_t^{(1)}$, $W_t^{(2)}$ and $W_t^{(3)}$ are independent Wiener processes driving each reaction channel. A typical realisation of the model is provided in Figure~\ref{fig:fig6}. Note that as $t \to \infty$ the stationary distribution is a product of Dirac distributions, that is, a point mass at $\bvec{X}_\infty = [E_0+C_0,0,0,S_0+C_0+P_0]^\text{T}$ given $\bvec{X}_0 = [E_0,S_0,C_0,P_0]^\text{T}$. Therefore observations involving the transient behaviour are essential to recover information about the rate parameters. 

While some analytic progress on likelihood approximation can be made using moment closures~\citep{Schnoerr2017}, the second order reaction for the production of complexes, $C$, effectively renders the distribution of the forwards problem analytically intractable.

\subsubsection{Time-course data and inference problem definition}
We generate synthetic data using a single realisation of the Michaelis-Menten chemical Langevin SDE with initial condition $\bvec{X}_0 = [100,100,0,0]^{\text{T}}$ and kinetic rate parameters $k_1 = 1\times10^{-3}$, $k_2 = 5\times 10^{-3}$ and $k_3 = 1\times 10^{-2}$. Observations are taken at $n = 20$ uniformly spaced time points $t_1 = 5,\, t_2 = 10,\ldots, \,t_{20} = 100$. The observation process considers Gaussian noise applied to each chemical species copy number with a standard deviation of $\sigma_{\text{obs}} = 10$, that is, $\bvec{Y}_\text{obs}^{(i)} \sim \mathcal{N}(\bvec{X}_{t_i},\sigma_{\text{obs}}^2 \bvec{I})$ where $\bvec{I}$ is the $4 \times 4$ identity matrix. See Appendix~\ref{sec:obs_data} for the resulting data table.

We perform inference on all three rate parameters, $\paramvec = [k_1,k_2,k_3]^\text{T}$. We use the particle MCMC approach to sample the Bayesian posterior,
\begin{equation*}
\CondPDF{k_1,k_2,k_3}{\mathcal{D}} \propto \like{k_1,k_2,k_3}{\mathcal{D}}\PDF{k_1,k_2,k_3},
\end{equation*}
where $\PDF{k_1,k_2,k_3}$ is the joint uniform prior with independent components $k_1 \sim \mathcal{U}(0,5\times10^{-3})$, $k_2 \sim \mathcal{U}(0,2.5\times10^{-2})$ and $k_3 \sim \mathcal{U}(0,5\times10^{-2})$. The likelihood is estimated using the bootstrap particle filter (Algorithm~\ref{alg:BPF}) with $R = 100$ particles and the Euler-Maruyama method for simulation with $\Delta t = 0.1$.

\subsubsection{Chain initialisation and proposal tuning} 

Any application of MCMC requires both a method of initialising the chain and choosing the proposal kernel. To deal with both of these challenges we can apply trial chains. 

\begin{landscape}
	\begin{figure}
		\centering
		\includegraphics[width=1\linewidth]{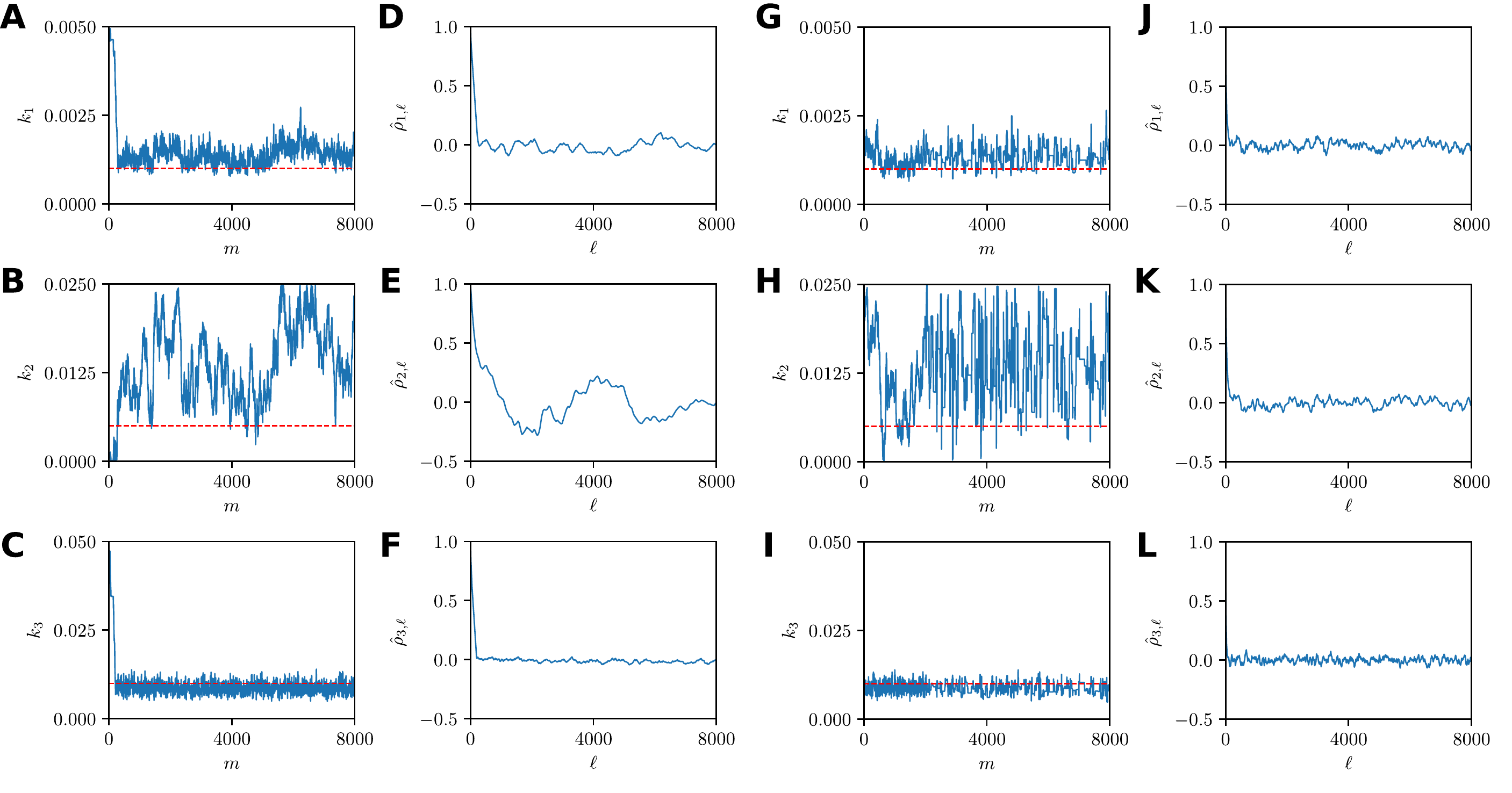}
		\caption{Comparison of marginal trace plots and autocorrelation functions (see Appendix~\ref{sec:app_conv_diag}) using (A)-(F) the na\"{i}ve independent Gaussian proposals and (G)-(L) optimally scaled correlated proposals.}
		\label{fig:fig7}
	\end{figure}
	
\end{landscape}

Four trial chains are simulated for $\mathcal{M} = 8,000$ iterations, each initialised with a random sample from the prior with a non-zero likelihood estimate. The proposal kernel used in all four trial chains is a Gaussian with covariance  matrix
\begin{equation*}
\boldsymbol{\Sigma} = \begin{bmatrix}
5.208\times 10^{-9} & 0 &0 \\
0  & 1.302 \times 10^{-7}  & 0 \\
0 & 0 & 5.208\times 10 ^{-7}
\end{bmatrix}.
\end{equation*}
The diagonal entries correspond to a proposal density such that one tenth of the prior standard deviation for each parameter is within a single standard deviation of the proposal; such independent proposal kernels are typical choices.
However, this proposal is not very efficient, as Figure~\ref{fig:fig7}(A)--(F) indicates for the first 8,000 iterations of the first chain. However, these chains are not used for inference, just for configuring a new set of more efficient chains.

The tuned proposal kernel is constructed by taking the total covariance matrix using the pooled sample of the four trial chains (total of $32,000$ samples),
\begin{equation*}
\hat{\boldsymbol{\Sigma}} = \begin{bmatrix}
2.693 \times 10 ^{-7} & 9.754 \times 10 ^{-7} & 1.536 \times 10 ^{-7}\\
9.754 \times 10 ^{-7} & 3.348\times 10 ^{-5} & -1.289 \times 10^{-5}\\
1.536 \times 10 ^{-7} &  -1.289 \times 10^{-5} & 4.677 \times 10^{-5}
\end{bmatrix},
\end{equation*}
and applying the optimal scaling rule from~\citet{Roberts2001}
\begin{equation*}
\boldsymbol{\Sigma}_{\text{opt}} = \frac{2.38^2}{3}\hat{\boldsymbol{\Sigma}}.
\end{equation*}
While the optimality of this scaling factor assumes a Gaussian posterior density, this is a useful guide that is widely applied~\citep{Roberts2009}. The final iteration of the trial chains is then used to initialise four new tuned chains with this optimal proposal covariance. The improvement in convergence behaviour is shown in Figure~\ref{fig:fig7}(G)--(L). 
\FloatBarrier

\subsubsection{Convergence assessment and parameter estimates}

Determining the number of iterations from the tuned chains to ensure valid inference is another practical challenge. Here, we follow the recommendations of~\cite{Vehtari2019} and apply the rank normalised $\hat{R}$ statistic along with the multiple chain effective sample size $S_\text{eff}$ (Appendix~\ref{sec:app_conv_diag}) using the four tuned Markov chains. Informally, $\hat{R}$ represents the ratio between an estimate of the posterior variance to the average variance of each independent Markov chain, and as $\mathcal{M} \to \infty$ then $\hat{R} \to 1$~\citep{Gelman1992}. The $S_\text{eff}$ statistic provides a measure of effective number of i.i.d. samples that the Markov chains represent for the purposes of computing an expectation. Larger values of $S_\text{eff}$ are better, but $S_{\text{eff}}$ will typically be much smaller than $\mathcal{M}$. 

The results, by parameter, are shown in Table~\ref{tab:michmentconv} after $\mathcal{M} = 15,000$ iterations per chain. \cite{Vehtari2019} recommend that $\hat{R} < 1.01$ and $S_\text{eff} > 400$ for each parameter. We conclude that the chains have converged sufficiently for our purposes.
\begin{table}[h]
	\centering
	\caption{Convergence diagnostics using four chains each with 15,000 iterations using the optimal proposal with dependent components. }	
	\begin{tabular}{r|ccc}
		& $k_1$ & $k_2$ & $k_3$\\
		\hline
		$S_\text{eff}$ & 986 & 683 & 1,909 \\
		$\hat{R}$ & 1.0046 & 1.0044 & 1.0023 
	\end{tabular}
	\label{tab:michmentconv}
\end{table}

\begin{figure}[h]
	\centering
	\includegraphics[width=1\linewidth]{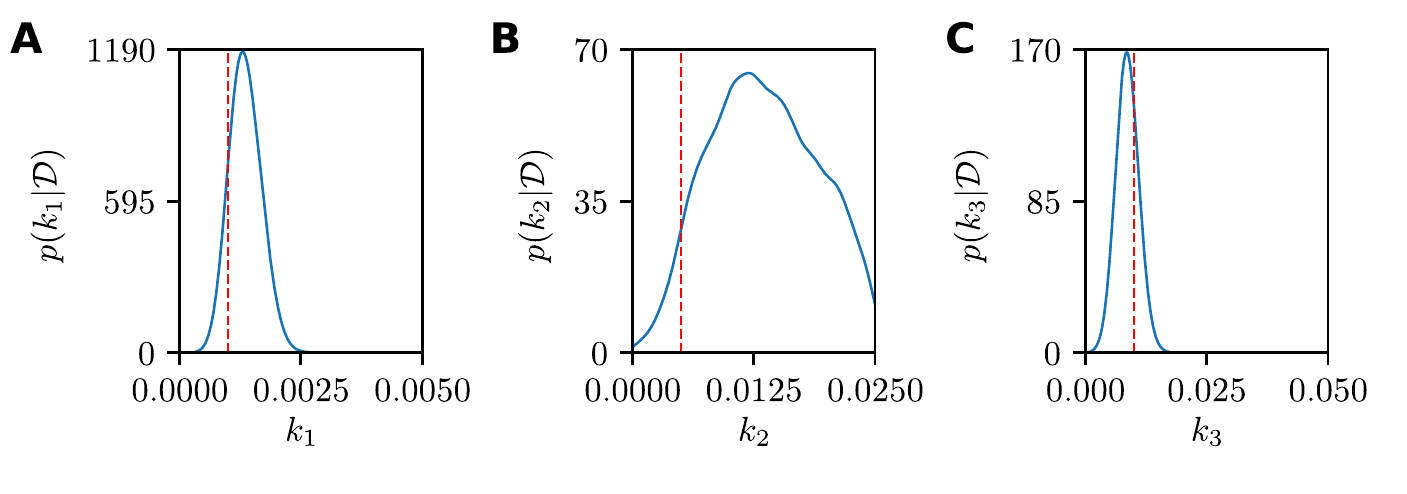}
	\caption{Marginal smoothed kernel density estimates for the Michaelis-Menten model using four converged particle MCMC chains. True parameter values are also indicated (red dashed).}
	\label{fig:fig8}
\end{figure}

The resulting inferences are shown in Table~\ref{tab:michmentinf} and Figure~\ref{fig:fig8}. For all parameters, the true values are within the range of the estimates obtained in 
Table~\ref{tab:michmentinf}. The marginal posterior densities shown in Figure~\ref{fig:fig8}. 

\begin{table}[h]
	\centering
	\caption{Parameter estimates based on estimates of the mean, $\hat{\mu}$, and standard deviation, $\hat{\sigma}$, with respect to the marginal posterior. }	
	\begin{tabular}{r|ccc}
		& $k_1$ & $k_2$ & $k_3$\\
		\hline
		$\paramvec_{\text{true}}$ & $1.000\times10^{-3}$ & $5.000\times10^{-3}$ & $1.000\times10^{-2}$ \\
		$\hat{\mu}$ & $1.365\times10^{-3}$ & $1.381\times10^{-2}$ & $8.640\times10^{-3}$ \\
		$\hat{\sigma}$ & $2.783\times10^{-4}$ & $5.441\times10^{-3}$ & $1.441\times10^{-3}$ 
	\end{tabular}
	\label{tab:michmentinf}
\end{table}

In Figure~\ref{fig:fig8}, we see that the modes of the marginal posteriors for both $k_1$ and $k_3$ are very close to the true values. However, for $k_2$ the mode overestimates the true parameter. It is important to emphasize, that this is not due to inaccuracy of the pseudo-marginal inference, but is a feature of the true posterior density. This result effectively highlights the uncertainty in the $k_2$ estimate due to partial observations, observation error, and model stochasticity.

The implementation of this inference problem, including data generation,  tuning and initialisation steps, convergence assessment, and plotting is given in \texttt{DemoMichMentPMCMC.jl}. The rank normalised $\hat{R}$ and $S_\text{eff}$ statistics are implemented within \texttt{Diagnostics.jl}. 
\FloatBarrier

\subsection{Example 2: The Schl\"{o}gl model}
The second example demonstrates the phenomenon of stochastic bi-stability. This leads to a very challenging inference problem that is poorly suited to alternative likelihood free schemes such as ABC. 

\subsubsection{Model definition}
\begin{figure}[h]
	\centering
	\includegraphics[width=1\linewidth]{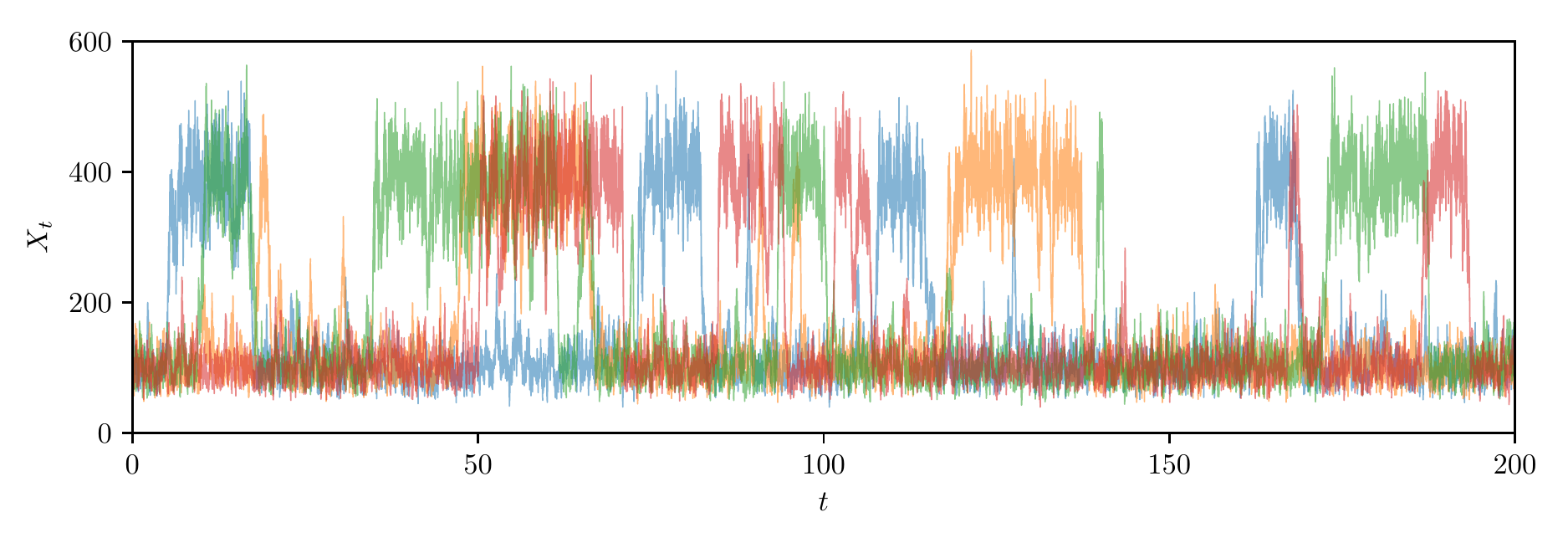}
	\caption{Four example realisations of the Schl\"{o}gl model demonstrating stochastic bi-stability. The simulations are produced using the Euler-Maruyama scheme with $\Delta t = 10^{-3}$, initial condition $X_0 = 0$, and parameters $k_1 = 1.8\times10^{-1}$, $k_2 = 2.5\times10^{-4}$, $k_3 = 2.2\times10^{3}$ and $k_4 = 3.75\times10^{1}$.}
	\label{fig:fig9}
\end{figure}

This example is a theoretical biochemical network initially studied by \cite{Schlogl1972}. This model involves a single chemical species, $X_t$, that evolves according to four reactions
\begin{equation}
2X \overset{k_1}{\rightarrow} 3X,\quad 3X \overset{k_2}{\rightarrow} 2X, \quad \emptyset \overset{k_3}{\rightarrow} X, \quad\text{and}\quad X \overset{k_4}{\rightarrow} \emptyset, \label{eq:schlogl}
\end{equation}
with propensities $a_1(X_t) = k_1X_t(X_t-1)$, $a_2(X_t) = k_2 X_t(X_t-1)(X_t-2)$, $a_3(X_t) = k_3$ and $a_4(X_t) = k_4X_t$
and stoichiometries $\nu_1 = 1$, $\nu_2 = -1$, $\nu_3 = 1$ and $\nu_4 = -1$. The Chemical Langevin It\={o} SDE is 
\begin{equation}
\begin{split}
\text{d}X_t =&\, [-k_2X_t^3 + (k_1+3k_2)X_t^2 - (k_1 + 2k_2 + k_4)X_t + k_3]\text{d}t \\ &+ \sqrt{k_2X_t^3 + (k_1 - 3k_2)X_t^2 + (2k_2 - k_1 + k_4)X_t + k_2}\text{d}W_t,
\end{split}
\end{equation}
where $W_t$ is a Wiener process. For certain values of the rate parameters the underlying deterministic model has two stable steady states separated by an unstable steady state~\citep{Schlogl1972,Vellela2009}. In the stochastic case, it is possible for the intrinsic noise of the system to drive $X_t$ from around one stable state toward the other; resulting in switching behaviour  demonstrated in Figure~\ref{fig:fig9} called \textit{stochastic bi-stability}.

The time between switching events is also a random variable, and observations taken from a single realisation will be very difficult to match using simulated data in the ABC setting, therefore acceptance rates will be prohibitively low. 
On the other hand particle MCMC is ideally suited to this problem since we condition simulations on the observations, thereby only sampling from realisations that pass closely to the data.

\FloatBarrier
\subsubsection{Time-course data and inference problem definition}
We generate synthetic data using a single realisation of the Schl\"{o}gl model chemical Langevin SDE with initial condition $X_0 = 0$ and kinetic rate parameters $k_1 = 1.8 \times 10^{-1}$, $k_2 = 2.5 \times 10^{-4}$, $k_3 = 2.2 \times 10^3$ and $k_4 = 3.75 \times 10^{1}$. Observations are taken at $n = 16$ uniformly spaced time points $t_1 = 12.5,\, t_2 = 25,\ldots,\,t_{16} = 200$. The observation process is modelled by Gaussian noise applied to the chemical species copy number with a standard deviation of $\sigma_{\text{obs}} = 10$, that is, $Y_\text{obs}^{(i)} \sim \mathcal{N}(X_{t_i},\sigma_{\text{obs}}^2)$. See Appendix~\ref{sec:obs_data} for the resulting data table.

We perform inference on all four rate parameters, $\paramvec = [k_1,k_2,k_3,k_4]^\text{T}$. We use the particle MCMC approach to sample for the Bayesian posterior,
\begin{equation*}
\CondPDF{k_1,k_2,k_3,k_4}{\mathcal{D}} \propto \like{k_1,k_2,k_3,_4}{\mathcal{D}}\PDF{k_1,k_2,k_3,k_4},
\end{equation*}
where $\PDF{k_1,k_2,k_3,k_4}$ is the joint uniform prior with independent components $k_1 \sim \mathcal{U}(0,5.4\times10^{-1})$, $k_2 \sim \mathcal{U}(0,7.5\times10^{-4})$, $k_3 \sim \mathcal{U}(0,6.6\times10^{3})$, and $k_4 \sim \mathcal{U}(0,1.125\times10^{2})$. The likelihood is estimated using the bootstrap particle filter (Algorithm~\ref{alg:BPF}) with $R = 100$ particles and Euler-Maruyama for simulation with $\Delta t = 0.1$.

\subsubsection{Chain initialisation and proposal tuning}

To initialise and tune four chains for inference on the four rate parameters of the Schl\"{o}gl model, we apply the same procedure as described for the Michaelis-Menten inference problem. The only difference is the number of samples applied.

\begin{landscape}
	\begin{figure}
		\centering
		\includegraphics[width=1\linewidth]{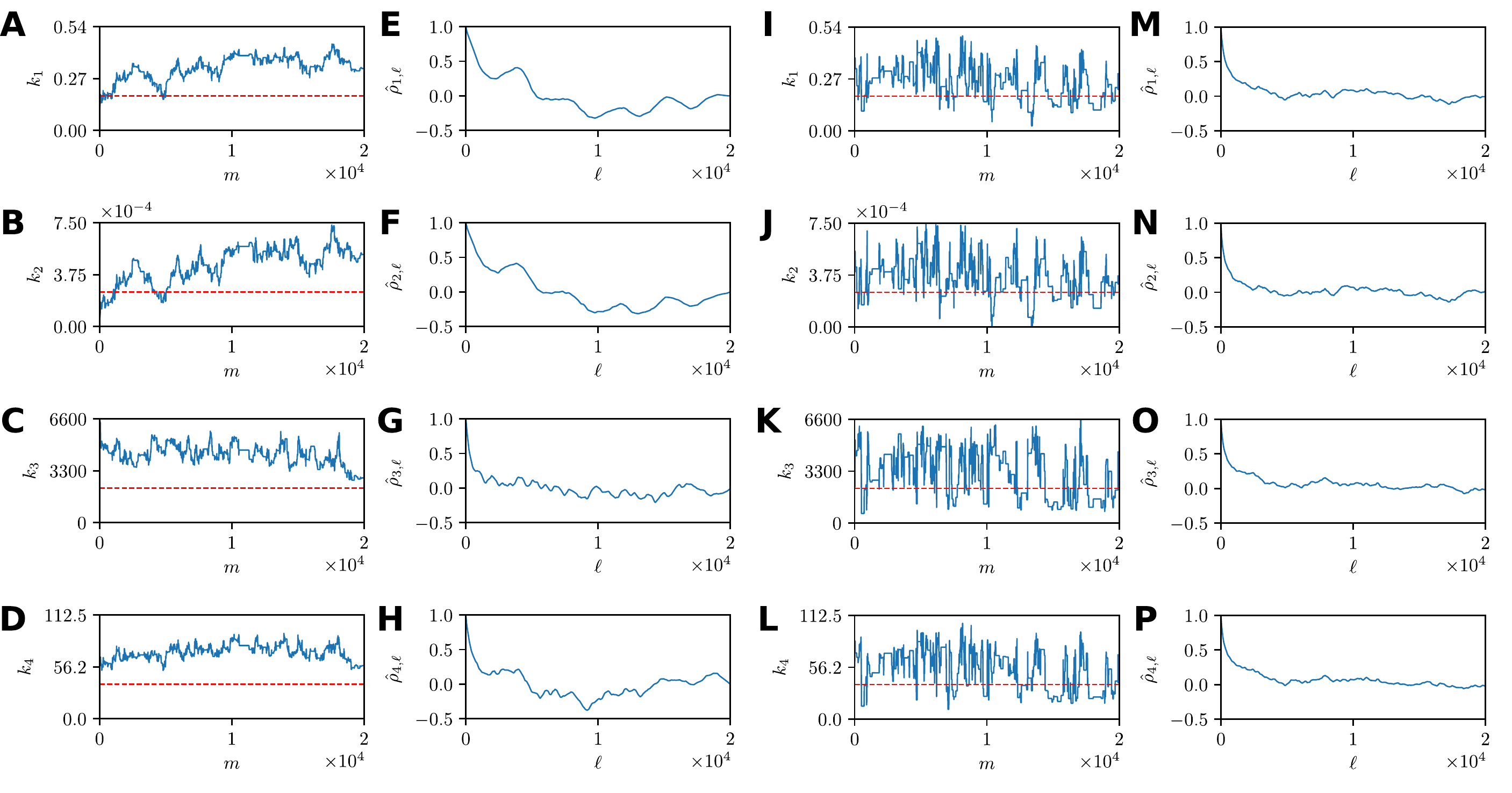}
		\caption{Comparison of marginal trace plots and autocorrelation functions (see Appendix~\ref{sec:app_conv_diag}) using (A)--(H) the na\"{i}ve independent Gaussian proposals and (I)-(P) optimally scaled correlated proposals.}
		\label{fig:fig10}
	\end{figure}
\end{landscape}

Firstly, four trial chains are simulated for $\mathcal{M} = 20,000$ iterations, each of these chains is initialised with a random sample from the prior with a non-zero likelihood estimate. The proposal kernel used in all four trial chains is Gaussian with covariance
\begin{equation*}
\boldsymbol{\Sigma} = \begin{bmatrix}
2.430 \times 10^{-4} & 0.0 & 0.0 & 0.0 \\
0.0 & 4.688 \times 10^{-10} & 0.0 & 0.0 \\
0.0 & 0.0 & 3.63\times 10^{4} & 0.0 \\
0.0 & 0.0 & 0.0 & 1.055 \times 10^{1}  
\end{bmatrix}.
\end{equation*}
Again, we start with a typical independent proposal kernel with diagonal entries calculated so that one proposal standard deviation in each parameter corresponds to one tenth the prior standard deviation. 
The tuned proposal kernel is constructed by taking the convariance of the pooled sample of the four trial chains (a total of $80,000$ samples),
\begin{equation*}
\hat{\boldsymbol{\Sigma}} = \begin{bmatrix}
4.770 \times 10^{-3} & 6.995 \times 10^{-4} & 4.429 \times 10^{1} & 9.297 \times 10^{-1}\\	
6.994 \times 10^{-6} & 1.511 \times 10^{-8} & 1.640 \times 10^{-3} & 5.929 \times 10^{-4} \\
4.429 \times 10^{1} & 1.640 \times 10^{-3} & 1.621 \times 10^{6} & 2.061 \times 10^{4} \\
9.297 \times 10^{-1} & 5.929 \times 10^{-4} & 2.061 \times 10^{4} & 3.199 \times 10^{2}
\end{bmatrix},
\end{equation*} 
and applying the optimal scaling rule from \cite{Roberts2001}
\begin{equation*}
\boldsymbol{\Sigma}_{\text{opt}} = \frac{2.38^2}{4} \hat{\boldsymbol{\Sigma}}.
\end{equation*}
The final iteration of the trial chains is then used to initialise for new chains using this optimal proposal covariance. Figure~\ref{fig:fig10} demonstrates the improvement in convergence behaviour. 

\subsubsection{Convergence assessment and parameter estimates}
Convergence diagnostic results, by parameter, are shown in Table~\ref{tab:schloglconv} after $\mathcal{M} = 240,000$ iterations per chain. Again, we ensure that the criteria of $\hat{R} < 1.01$ and $S_{\text{eff}} > 400$ \citep{Vehtari2019} are satisfied for all parameters. We note that the convergence rate of the MCMC chains is significantly slower than that of the Michaelis-Menten example. In practice, one might consider a fully adaptive proposal scheme for this model to improve convergence rates~\citep{Roberts2001,Roberts2009}. 
\begin{table}[h]
	\centering
	\caption{Convergence diagnostics using four chains each with $240,000$ iterations using the optimal proposal with dependent components. }	
	\begin{tabular}{r|cccc}
		& $k_1$ & $k_2$ & $k_3$ & $k_4$\\
		\hline
		$S_\text{eff}$ & 577 & 656& 467 & 625 \\
		$\hat{R}$ & 1.0054 & 1.0060 & 1.0049 & 1.0039
	\end{tabular}
	\label{tab:schloglconv}
\end{table}

\begin{figure}[h]
	\centering
	\includegraphics[width=0.8\linewidth]{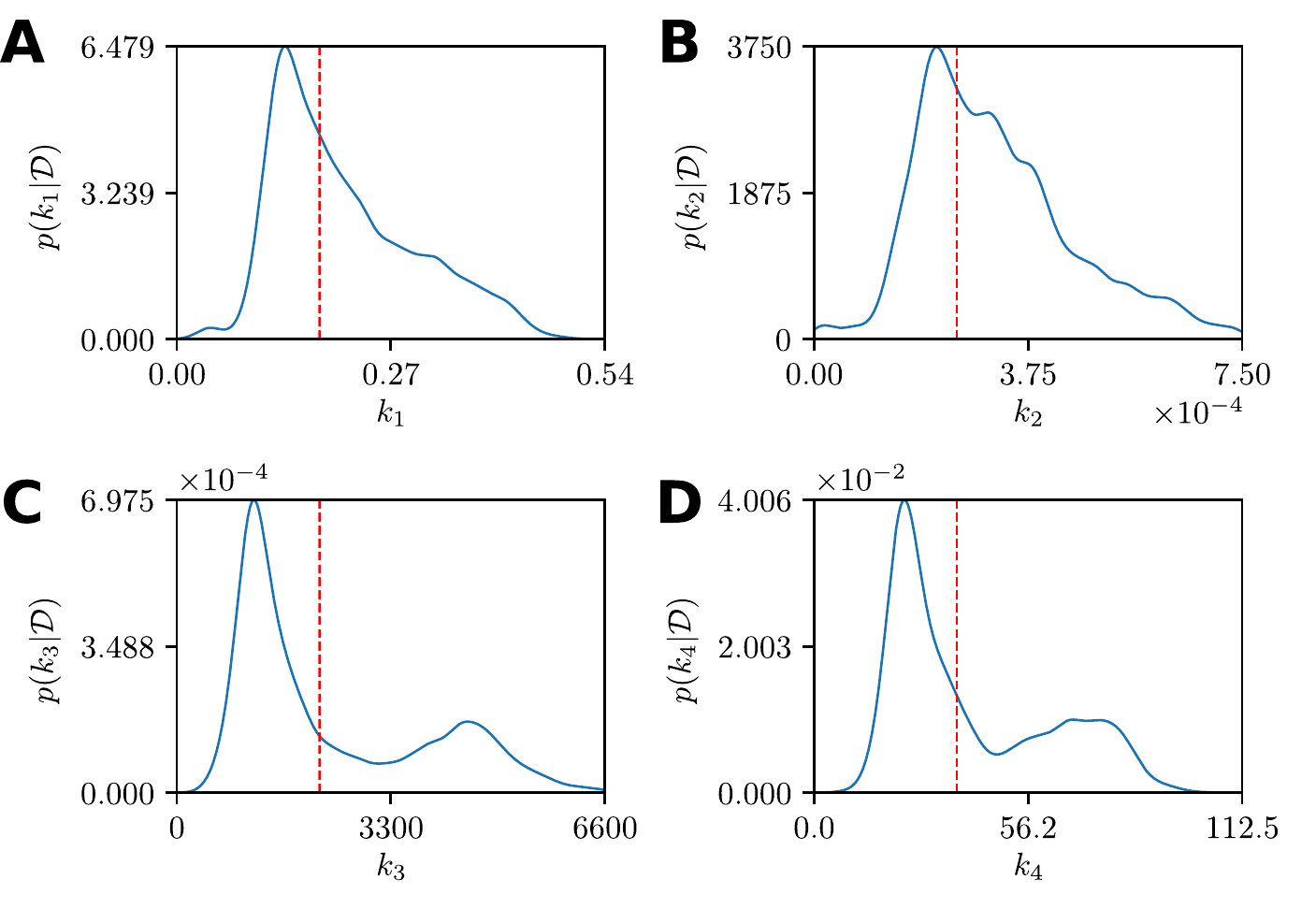}
	\caption{Marginal smoothed kernel density estimates for the Schl\"{o}gl model using four converged particle MCMC chains. True parameter values are also indicated (red dashed).}
	\label{fig:fig11}
\end{figure}

The resulting inferences are shown in Table~\ref{tab:schloglinf} and Figure~\ref{fig:fig11}. For all parameters, the true values are within the range of the estimates obtained in 
Table~\ref{tab:schloglinf}. The marginal posterior densities shown in Figure~\ref{fig:fig11}. 

Figure~\ref{fig:fig11}(A)--(B) demonstrates that both the posterior modes for $k_1$ and $k_2$ are very close to the true parameter values, however the uncertainties are asymmetric, indicating a range of possibly appropriate parameter values greater than the true values. The posteriors of $k_3$ and $k_4$ are very interesting as they are bi-modal (Figure~\ref{fig:fig11}(C)--(D)). While the higher density model is closer to the true parameter values, the second lower density mode indicates that an alternative parameter combination in $k_3$ and $k_4$ can lead to very similar stochastic bi-stability in the Schl\"{o}gl model evolution. To observe this posterior bi-modality using ABC methods would be very challenging since $\epsilon$ would need to be prohibitively small. Furthermore, most expositions on ABC methods~\citep{Sunnaker2013,Toni2008,Warne2019} do not deal with multimodal posteriors.

\begin{table}[h]
	\centering
	\caption{Parameter estimates based on estimates of the mean, $\hat{\mu}$, and standard deviation, $\hat{\sigma}$, with respect to the marginal posterior. }	
	\begin{tabular}{r|cccc}
		& $k_1$ & $k_2$ & $k_3$ & $k_4$\\
		\hline
		$\paramvec_{\text{true}}$ & $1.800\times10^{-1}$ & $2.500\times10^{-4}$ & $2.200\times10^{3}$ & $3.750\times 10^{1}$ \\
		$\hat{\mu}$ & $2.118\times10^{-1}$ & $3.200\times10^{-4}$ & $2.350\times10^{3}$ & $4.104\times 10^{1}$\\
		$\hat{\sigma}$ & $8.988\times10^{-2}$ & $1.377\times10^{-4}$ & $1.483\times10^{3}$ & $2.121 \times 10^{1}$ 
	\end{tabular}
	\label{tab:schloglinf}
\end{table}

The apparent bi-modal nature of parameters $k_3$ and $k_4$ (Figure~\ref{fig:fig11}) could be a reason for the increased computational requirements of this model, since all chains must occupy both modes sufficiently to reduce $\hat{R}$ and to increase $S_\text{eff}$ sufficiently.   

The implementation of this inference problem, including data generation,  tuning and initialisation steps, convergence assessment, and plotting is give in \texttt{DemoSchloglPMCMC.jl}. The rank normalised $\hat{R}$ and $S_\text{eff}$ statistics are implemented within \texttt{Diagnostics.jl}.

\FloatBarrier

\subsection{Example 3: The repressilator model}
The last example we consider in this work is a gene regulatory network, originally realised synthetically by~\cite{Elowitz2000}, that includes a feedback loop resulting in stochastic oscillatory dynamics in the gene expression. The model is of interest in biological studies~\citep{Pokhilko2012,PotvinTrottier2016} and is a challenging benchmark for inference methods~\citep{Toni2008}.

\subsubsection{Model definition}
\begin{figure}[h]
	\centering
	\includegraphics[width=1\linewidth]{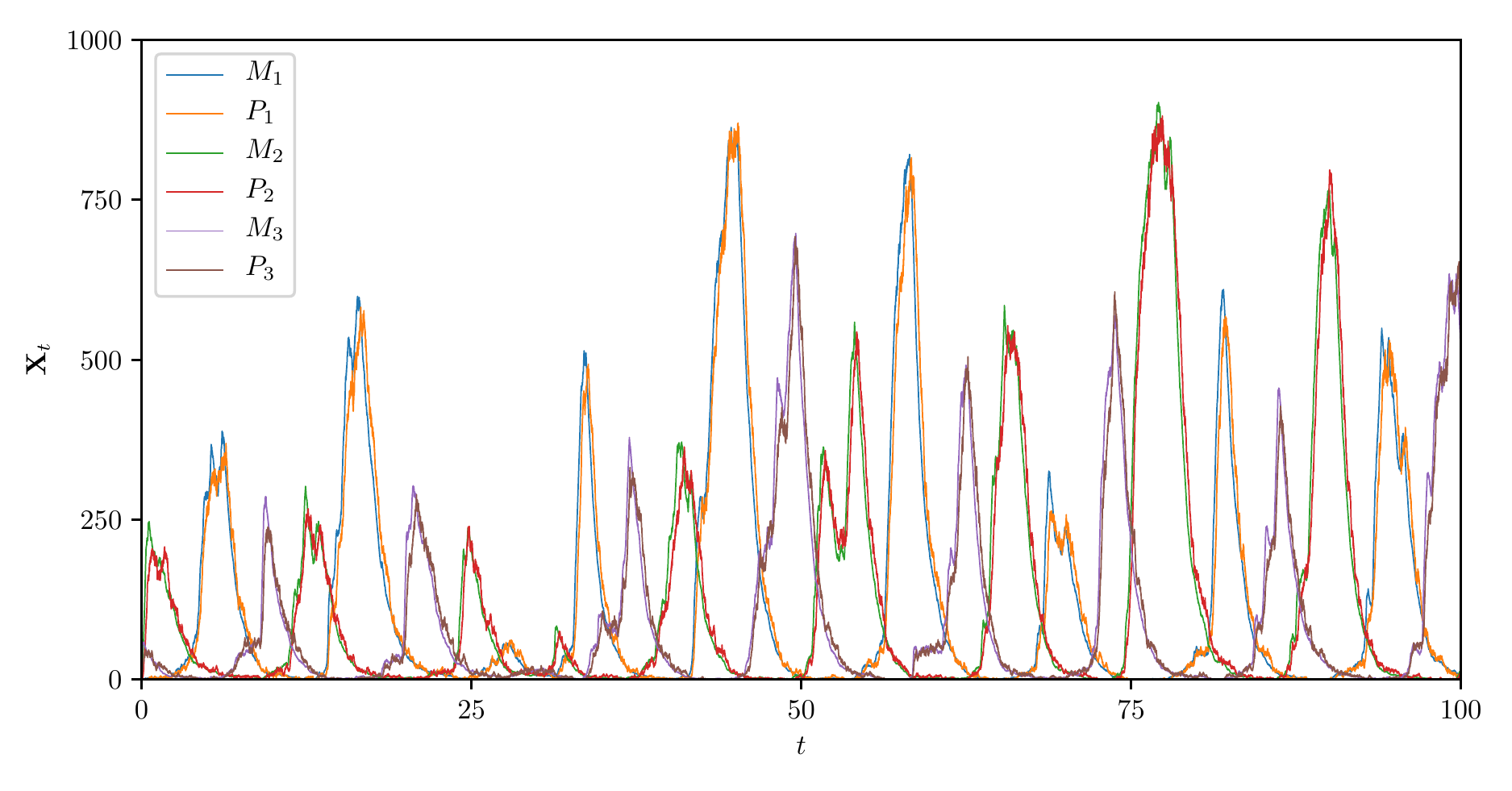}
	\caption{Example realisation of the repressilator model demonstrating oscillatory gene expression. The simulation is produced using the Euler-Maruyama scheme with $\Delta t = 1\times10^{-3}$, initial condition $\bvec{X}_0 = [M_{1,0},P_{1,0},M_{2,0},P_{2,0},M_{3,0},P_{3,0}]^\text{T} = [0,2,0,1,0,3]^\text{T}$, and parameters $\alpha_0 = 1$, $\alpha = 1000$, $n = 2$, $\beta = 5$, and $\gamma = 1$.}
	\label{fig:fig12}
\end{figure}

The repressilator consists of three genes where the expression of one gene inhibits the expression of the next gene, forming a feedback loop between the three genes. The regulatory network consists of twelve reactions describing the transcription of the three mRNAs, $M_1, M_2,$ and $ M_3$, associated with each gene, $G_1, G_2,$ and $G_3$, their expression through translation into proteins, $P_1, P_2,$ and $ P_3$, and degradation processes for both mRNAs and proteins. For the $i$th gene we have,
\begin{equation}
\underbrace{G_i \xlongrightarrow{\alpha_0 + \alpha/(1+P_j^n)} M_i}_{\substack{\text{mRNA transcription}}},\quad  \underbrace{M_i \overset{\beta}{\rightarrow} M_i + P_i}_{\substack{\text{protein translation}}}, \quad  \underbrace{P_i \overset{\beta}{\rightarrow} \emptyset}_{\substack{\text{protein degradation}}}, \quad\text{and}\quad\underbrace{M_i \overset{\gamma}{\rightarrow} \emptyset}_{\substack{\text{mRNA degradation}}}, \label{eq:repr}
\end{equation}
where $j = (i+1 \mod 3) +1$, $\alpha_0 \geq 0$ is the leakage transcription rate (transcription rate of maximally inhibited gene), $\alpha + \alpha_0 > 0$ is the free transcription rate (uninhibited transcription rate), $n \geq 0$ is the Hill coefficient that describes the strength of the repressive effect of the inhibitor protein $P_j$, $\beta > 0$ is the protein translation and degradation rate, and $\gamma > 0$ is the mRNA degradation rate~\citep{Elowitz2000}. The gene copy numbers are fixed at $G_i = 1$ for $i = 1, 2, 3$. The resulting chemical Langevin approximation (\eqref{eq:CLE}) to the repressilator model (\eqref{eq:repr})  leads to a coupled system of It\={o} SDEs
\begin{equation}
\label{eq:CLErepr}
\begin{split}
\text{d}M_{1,t} &= \left(\alpha_0 + \frac{\alpha}{1+P_{3,t}^n} - \gamma M_{1,t}\right)\text{d}t + \sqrt{\alpha_0 + \frac{\alpha}{1+P_{3,t}^n}}\text{d}W_t^{(1)} - \sqrt{\gamma M_{1,t}}\text{d}W_t^{(4)},\\
\text{d}P_{1,t} &= \beta\left(M_{1,t} - P_{1,t}\right)\text{d}t + \sqrt{\beta M_{1,t}}\text{d}W_t^{(2)} - \sqrt{\beta P_{1,t}}\text{d}W_t^{(3)}, \\
\text{d}M_{2,t} &= \left(\alpha_0 + \frac{\alpha}{1+P_{1,t}^n} - \gamma M_{2,t}\right)\text{d}t + \sqrt{\alpha_0 + \frac{\alpha}{1+P_{1,t}^n}}\text{d}W_t^{(5)} - \sqrt{\gamma M_{2,t}}\text{d}W_t^{(8)},\\
\text{d}P_{2,t} &= \beta\left(M_{2,t} - P_{2,t}\right)\text{d}t + \sqrt{\beta M_{2,t}}\text{d}W_t^{(6)} - \sqrt{\beta P_{2,t}}\text{d}W_t^{(7)}, \\
\text{d}M_{3,t} &= \left(\alpha_0 + \frac{\alpha}{1+P_{2,t}^n} - \gamma M_{3,t}\right)\text{d}t + \sqrt{\alpha_0 + \frac{\alpha}{1+P_{2,t}^n}}\text{d}W_t^{(9)} - \sqrt{\gamma M_{3,t}}\text{d}W_t^{(12)},\\
\text{d}P_{3,t} &= \beta\left(M_{3,t} - P_{3,t}\right)\text{d}t + \sqrt{\beta M_{3,t}}\text{d}W_t^{(10)} - \sqrt{\beta P_{3,t}}\text{d}W_t^{(11)},
\end{split}
\end{equation}
where $W_t^{(1)},W_t^{(2)}, \ldots, W_t^{(12)}$ are independent Wiener processes driving each reaction channel. Certain parameter combinations lead to stochastic oscillations in the gene expression levels, that is, the protein copy numbers associated with the expressed gene. Figure~\ref{fig:fig12} demonstrates this behaviour in which the expressed gene alternates between $G_2$, $G_1$, and $G_3$ in sequence due to the feedback loop in the gene inhibitor network.  

\FloatBarrier

\subsubsection{Time-course data and inference problem definition}
We generate synthetic data using a single realisation of the repressilator chemical Langevin SDE with initial condition $\bvec{X}_0 = [M_{1,0},P_{1,0},M_{2,0},P_{2,0},M_{3,0},P_{3,0}]^\text{T} = [0,2,0,1,0,3]^\text{T}$ and parameters $\alpha_0 = 1$, $\alpha = 1000$, $n = 2$, $\beta = 5$, and $\gamma = 1$. Observations are taken at $n = 20$ uniformly spaced time points $t_1 = 5$, $t_2 = 10$, $\ldots, t_{20} = 100$. Again we consider Gaussian noise applied to each chemical species copy number with a standard deviation of $\sigma_{\text{obs}}=10$, that is, $\bvec{Y}_{\text{obs}}^{(i)} \sim \mathcal{N}\left(\bvec{X}_{t_i},\sigma_{\text{obs}}^2\bvec{I}\right)$ where $\bvec{I}$ is the $6 \times 6$ identity matrix. See Appendix~\ref{sec:obs_data} for the resulting data table.

We perform  inference on four of the model parameters, $\paramvec = [\alpha_0,\alpha,n,\beta]^{\text{T}}$, and fix the mRNA degradation rate $\gamma = 1$. We use the particle MCMC approach to sample from the Bayesian posterior,
\begin{equation*}
\CondPDF{\alpha_0,\alpha,n,\beta}{\dat} \propto \like{\alpha_0,\alpha,n,\beta}{\dat}\PDF{\alpha_0,\alpha,n,\beta},
\end{equation*}
where $\PDF{\alpha_0,\alpha,n,\beta}$ is the joint uniform prior with independent components $\alpha_0\sim \mathcal{U}(0,10)$, $\alpha \sim \mathcal{U}(500,2500)$, $n \sim \mathcal{U}(0,10)$, and $\beta \sim \mathcal{U}(0,20)$. 

\subsubsection{Chain initialisation and proposal tuning}

The same initialisation and proposal tuning procedure applied to the Michaelis-Menten and Schl\"{o}gl models is applied here. The resulting tuned proposal kernel convariance is given by
\begin{equation*}
\boldsymbol{\Sigma}_\text{opt} = \frac{2.38^2}{4}\begin{bmatrix}
43634.288  & 95.328 & 173.584 & 416.368 \\
95.328 & 0.345 & 1.051 & 1.142 \\
173.584 & 1.051 & 5.286 & 3.488 \\
416.368 & 1.142 & 3.488 & 5.939  
\end{bmatrix},
\end{equation*}
which is derived through application of the \cite{Roberts2001} scaling rule to the covariance matrix of the pooled samples from four trial chains, each with $\mathcal{M} = 5,000$ iterations. The trial chains are initialised and constructed in the same way as for the Michaelis-Menten and Schl\"{o}gl models.

\begin{figure}[h]
	\centering
	\includegraphics[width=0.6\linewidth]{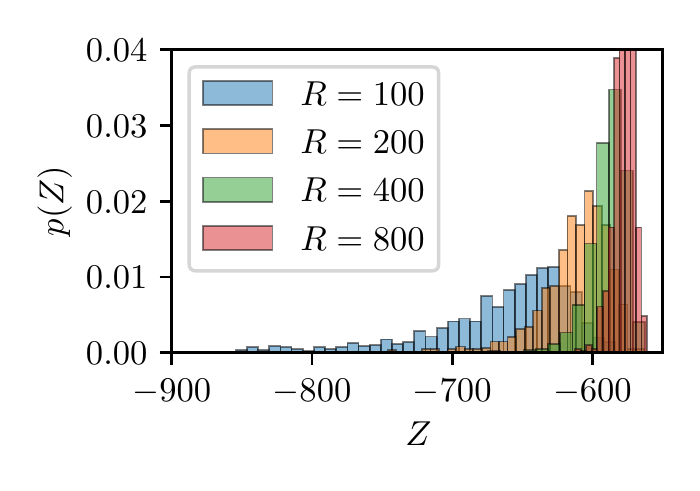}
	\caption{Distribution of $1,000$ likelihood estimates around a high density posterior point for different numbers of particles in the bootstrap particle filter. Significant bias is introduced for the lower particle counts.}
	\label{fig:reprll}
\end{figure}

The repressilator model is a good example of when one must be careful to use a large enough number of particles in the bootstrap particle filter.
Unlike the Michaelis-Menten and Schl\"{o}gl models, the repressilator model likelihood estimator is highly variable for low particle numbers.
Figure~\ref{fig:reprll} demonstrates the effect of the number of particles, $R$, on the distribution of the logarithm of the likelihood estimator evaluated $Z = \log \mclike{\paramvec}{\dat}$. 

Note that as $R$ decreases, not only does the variance of the estimator increase, but so does the bias that is seen through the shift in the estimator mode. Here, there is a trade-off, $R = 800$ yields a low variance and is much closer to the optimal criterion of~\cite{Doucet2015}. However, $ R = 400$ has a very similar mode, but slightly higher variance. 
This motivates the use of $R = 400$ particles to achieve reasonable convergences rates without too much additional computational burden.

\subsubsection{Convergence assessment and parameter estimates}

Convergence diagnostic results, by parameter, are shown in Table~\ref{tab:reprconv} after $\mathcal{M} = 145,000$ iterations per chain. In this case, the conservative convergence criteria of~\cite{Vehtari2019} have not yet been met. We report the results without additional computational effort for the purposes of this review, but we emphasise that for a real application more iterations of the MCMC chains should be performed to have confidence in the final inferences. Furthermore, it is important to note that $\hat{R} < 1.1$ is still a widely used convergence criterion~\citep{Gelman1992,Gelman2014}.

\begin{table}[h]
	\centering
	\caption{Convergence diagnostics using four chains each with $145,000$ iterations using the optimal proposal with dependent components.}	
	\begin{tabular}{r|cccc}
		& $\alpha$ & $\alpha_0$ & $\beta$ & $n$\\
		\hline
		$S_\text{eff}$ & 102 & 107& 152 & 205 \\
		$\hat{R}$ & 1.0358 & 1.0455 & 1.0152 & 1.0184
	\end{tabular}
	\label{tab:reprconv}
\end{table}

\begin{figure}[h]
	\centering
	\includegraphics[width=0.8\linewidth]{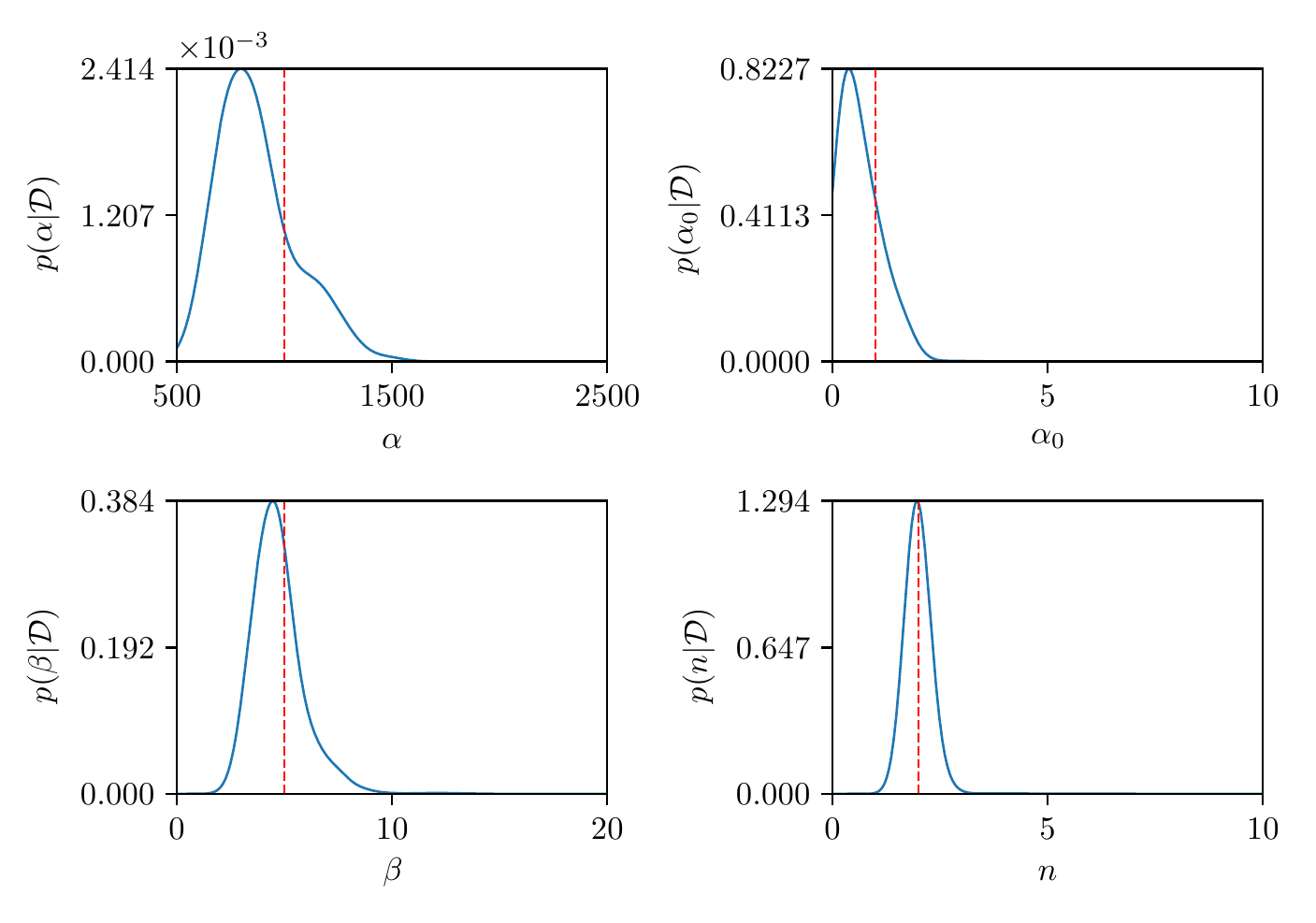}
	\caption{Marginal smoothed kernel density estimates for the repressilator model using four particle MCMC chains. True parameter values are also indicated (red dashed).}
	\label{fig:fig13}
\end{figure}

The resulting inferences are shown in Figure~\ref{fig:fig13} and Table~\ref{tab:reprinf}. For all parameters, the true values are within range of the estimates obtained in Table~\ref{tab:reprinf}. The marginal posterior densities are shown in Figure~\ref{fig:fig13}.

\begin{table}[h]
	\centering
	\caption{Parameter estimates based on estimates of the mean, $\hat{\mu}$, and standard deviation, $\hat{\sigma}$, with respect to the marginal posterior.}	
	\begin{tabular}{r|cccc}
		& $\alpha$ & $\alpha_0$ & $\beta$ & $n$\\
		\hline
		$\paramvec_{\text{true}}$ & $1000.0000$ & $1.0000$ & $5.0000$ & $2.0000$ \\
		$\hat{\mu}$ & $871.4327$ & $0.6224$ & $4.7106$ & $1.9960$\\
		$\hat{\sigma}$ & $172.9650$ & $0.4685$ & $1.1671$ & $0.2553$ 
	\end{tabular}
	\label{tab:reprinf}
\end{table}

For all parameters, the marginal posterior densities are uni-model, with modes that are close to the true parameter estimates. In particular, the $\beta$ and $n$ parameters are very accurately retrieved, whereas the marginal posteriors for $\alpha$ and $\alpha_0$ lead to underestimates. However, these underestimates are consistent with previous results~\citep{Toni2008}. A likely cause of this is the temporal sparsity of observations, leading to few observations of the peak gene expression levels (see Figure~\ref{fig:fig12} and data in Appendix~\ref{sec:obs_data}), as $\alpha$ and $\alpha_0$ relate to the transcription rate of the mRNAs. Despite the additional computational complexity associated with the particle filter for this inference problem, the repressilator model provides an insightful example of the efficacy of the pseudo-marginal approach to resolve biological parameters associated with gene regulation using synthetic data that is biological realisable.

The implementation of this inference problem, including data generation,  tuning and initialisation steps, convergence assessment, and plotting is give in \texttt{DemoRepressilatorPMCMC.jl}. The rank normalised $\hat{R}$ and $S_\text{eff}$ statistics are implemented within \texttt{Diagnostics.jl}.

\section{Summary}

In this work, we provide a practical guide to computational Bayesian inference using the pseudo-marginal approach~\citep{Andrieu2009,Beaumont2003,Andrieu2010}. We compare and contrast, using a tractable example, the pseudo-marginal approach with the ABC alternative ~\citep{Sisson2018,Sunnaker2013}. Throughout, chemical Langevin SDE descriptions of biochemical reaction networks~\citep{Gillespie2000,Higham2008}, of various degrees of complexity, have been employed to demonstrate practical considerations when using these techniques to inference problems with intractable likelihoods.

The ABC approach to likelihood-free inference is widely applicable and used extensively in practical applications~\citep{Browning2018,Johnston2016,Kursawe2018,Warne2019b,Wilkinson2011}. For some applications, however, it can be difficult to determine \emph{a priori} an appropriate discrepancy metric and acceptance threshold for reliable inference. Furthermore, a sufficiently small threshold for the desired level of accuracy may result in prohibitively low acceptance rates~\citep{Sisson2007}. Pseudo-marginal methods do not suffer from these accuracy considerations since they converge to the true posterior target regardless of the variance of the estimator~\citep{Golightly2011}. As a result, the pseudo-marginal approach is significantly less sensitive to user-specified algorithm parameters than likelihood-free inference based on ABC.

There are also disadvantages to the pseudo-marginal approach. Firstly, it is not as generally applicable as ABC; the pseudo-marginal method requires an unbiased estimator, whereas ABC  only needs a model simulation process. While convergence to the true posterior distribution is not affected by the estimator variance, the rate of convergence is~\citep{Andrieu2009}; to obtain optimal likelihood variances, a large number of particles may be required, thus evaluating the likelihood estimate will be very expensive. Alternatively, ABC will only ever use a single simulation per iteration. Furthermore, under the assumption of observation error and model miss-specification, convergence to the true posterior is not always a significant advantage~\citep{Wilkinson2009,Andrieu2018} and ABC may be effectively considered exact~\citep{Wilkinson2013}. Lastly, pseudo-marginal methods are not widespread in the systems biology literature and there is a lack of exemplars, despite their suitability for many problems of interest. This review is intended to address this by presenting all the steps involved clearly and providing user-friendly implementations in an open access environment (\href{https://github.com/davidwarne/Warne2019_GuideToPseudoMarginal}{https://github.com/davidwarne/Warne2019\_GuideToPseudoMarginal}).  

For practical illustrative purposes, we focus on the fundamental method of particle marginal Metropolis-Hastings~\citep{Andrieu2010} using the bootstrap particle filter~\citep{Gordon1993} for likelihood estimation. There are many other variants to this classic approach, such as particle Gibbs sampling~\citep{Andrieu2010,Doucet2015}, coupled Markov chains~\citep{Dodwell2015,Dodwell2019}, and more advanced particle filters~\citep{Doucet2011} and proposal mechanisms~\citep{Botha2019,Cotter2013}. It is also important to note that the pseudo-marginal approach is equally valid for Bayesian sampling strategies based on sequential Monte Carlo~\citep{DelMoral2006,Sisson2007,Li2019}. Furthermore, advances in stochastic simulation~\citep{Schnoerr2017,Warne2019} can also improve the performance of the likelihood estimator, and the application of multilevel Monte Carlo to particle filters can further reduce estimator variance~\citep{Jasra2017,Jasra2018,Gregory2016}. 

Likelihood-free methods are essential to modern biological sciences, since many mechanistic models of interest have intractable likelihoods. Unlike ABC methods, the pseudo-marginal approach does not affect the stationary distribution for the purposes of MCMC sampling; this is a desirable property. However, one reason for the popularity and success of ABC methods has been its simplicity to implement. Through this accessible and practical demonstration, along with example open-source codes, the pseudo-marginal approach may become an additional readily available tool for likelihood-free inference within the wider scientific community.

\paragraph{Software availability}
The Julia code examples and demonstration scripts are available from GitHub \href{https://github.com/davidwarne/Warne2019_GuideToPseudoMarginal}{https://github.com/davidwarne/Warne2019\_GuideToPseudoMarginal}.
\paragraph{Acknowledgements}
This work was supported by the Australian Research Council (DP170100474). M.J.S. appreciates support from the University of Canterbury Erskine Fellowship. R.E.B. would like to thank the Leverhulme Trust for a Leverhulme Research Fellowship, the Royal Society for a Wolfson Research Merit Award, and the BBSRC for funding via BB/R00816/1.



\newpage
\begin{appendices}
	\appendix
	
	\section{Derivation of stationary distribution for the production-degradation model}
	\label{sec:app_stat_pd}
	\numberwithin{equation}{section}
	\numberwithin{figure}{section}
	\numberwithin{algorithm}{section}
	\numberwithin{table}{section}
	\setcounter{equation}{0}
	Here, we derive the solution to the stationary distribution for the production-degradation model. First, recall that the general form for the chemical Langevin equation is
	\begin{equation*}
		\text{d}\bvec{X}_t = \sum_{j=1}^M \boldsymbol{\nu}_ja_j(\bvec{X}_t)\text{d}t + \sum_{j=1}^M \boldsymbol{\nu}_j\sqrt{a_j(\bvec{X}_t)}\text{d}W_t^{(j)},
	\end{equation*}
	where $\bvec{X}_t$ takes values in $\mathbb{R}^N$, $W_t^{(1)},W_t^{(2)},\ldots,W_t^{(M)}$ are independent scalar Wiener processes, $\boldsymbol{\nu}_1,\boldsymbol{\nu}_2,\ldots,\boldsymbol{\nu}_M$ are the stoichiometric vectors and $a_1(\bvec{X}_t),a_2(\bvec{X}_t),\ldots,a_M(\bvec{X}_t)$ the propensity functions. Consider the distribution of $\bvec{X}_t$ over all possible realisations at time $t$ with probability density function $p(\bvec{x},t)$. The Fokker-Planck equation describes the forward evolution of this probability density in time. For the general chemical Langevin equation, the Fokker-Planck equation is given by
	\begin{equation}
		\pdydx{p(\bvec{x},t)}{t} =  \frac{1}{2}\sum_{j=1}^M \boldsymbol{\nu}_j^\text{T} \bvec{H}\left(a_j(\bvec{x})p(\bvec{x},t)\right)\boldsymbol{\nu}_j -\sum_{j=1}^M \nabla\left[a_j(\bvec{x})p(\bvec{x},t)\right]\boldsymbol{\nu}_j, \label{eq:FPE}
	\end{equation}
	where, for reaction $j$, $\nabla\left[a_j(\bvec{x})p(\bvec{x},t)\right]$ and $\bvec{H}\left(a_j(\bvec{x})p(\bvec{x},t)\right)$ are, respectively, the gradient vector and Hessian matrix of the product $a_j(\bvec{x})p(\bvec{x},t)$ with respect to the state vector $\bvec{x}$.
	
	For the production-degradation model, we have a single chemical species $X_t$, propensity functions
	\begin{equation}
		a_1(X_t) = k_1\quad \text{and}\quad a_2(X_t) = k_2X_t, \label{eq:proppd}
	\end{equation}
	with rate parameters $k_1$ and $k_2$, and stoichiometries
	\begin{equation} 
		\nu_1 = 1\quad \text{and}\quad \nu_2 = -1.\label{eq:stopd}
	\end{equation}
	By substituting \eqref{eq:proppd} and \eqref{eq:stopd} into \eqref{eq:FPE}, we obtain the Fokker-Planck equation for the production degradation model,
	\begin{equation}
		\pdydx{p(x,t)}{t} = \pddydx{}{x}\left[\frac{k_1 + k_2 x}{2}p(x,t)\right] - \pdydx{}{x}\left[(k_1 - k_2 x)p(x,t)\right]. \label{eq:FPEpd}
	\end{equation}
	The stationary distribution of $X_t$ corresponds to the steady state solution of \eqref{eq:FPEpd}, that is, $p_s(x) = \lim_{t \to \infty} p(x,t)$. The stationary probability density function, $p_s(x)$, satisfies
	\begin{equation}
		\ddydx{}{x}\left[\frac{k_1 + k_2 x}{2}p_s(x)\right] - \dydx{}{x}\left[(k_1 - k_2 x)p_s(x)\right] = 0. \label{eq:FPEstatpd}
	\end{equation}
	
	To obtain a solution, integrate~\eqref{eq:FPEstatpd} to obtain 
	\begin{equation}
		\dydx{p_s(x)}{x} + \left(\frac{k_2(1-x) -2k_1}{k_1 + k_2 x}\right)p_s(x) = C, \label{eq:sf}
	\end{equation}
	where $C$ is an arbitrary constant. We obtain $C = 0$ by assuming the boundary condition $\lim_{x \to \infty} p_s(x) = 0$. The solution to \eqref{eq:sf} can be obtained using an integrating factor,
	\begin{align}
		p_s(x) &= \frac{\tilde{C}}{k_1 + k_2x}\exp\left(2\int_{0}^{x}\frac{k_1-k_2y}{k_1+k_2 y}\,\text{d}y\right) \notag\\
		&= \tilde{C}\exp\left({-2x + \left(\frac{4k_1}{k_2}-1\right)\ln\left(k_1+k_2x\right)}\right), \label{eq:num} 
	\end{align}
	where $\tilde{C}$ is a constant that is obtained by enforcing the condition $\int_{-\infty}^\infty p_s(x)\,\text{d}x = 1$, yielding the stationary probability density provided in the main manuscript.
	
	\section{Pseudo-marginal MCMC as an exact approximation}
	\label{sec:app_pm_exact}
	Here we briefly explain why the stationary distribution of the pseudo-marginal Metropolis-Hastings method is the exact posterior. For more detailed analysis, see~\citet{Andrieu2009}, and \citet{Golightly2008}. Consider the following algebraic manipulations applied to the pseudo-marginal Metropolis-Hastings acceptance probability. We start with
	\begin{align*}
		\alpha(\paramvec^{*},\paramvec_{m}) &= \frac{\Kernel{\paramvec_{m}}{\paramvec^*}\mclike{\paramvec^{*}}{\dat}\PDF{\paramvec^*}}{\Kernel{\paramvec^{*}}{\paramvec_{m}}\mclike{\paramvec_{m}}{\dat}\PDF{\paramvec_{m}}} \\ &= \frac{\Kernel{\paramvec_{m}}{\paramvec^*}\like{\paramvec^{*}}{\dat}\left[\dfrac{\mclike{\paramvec^{*}}{\dat}}{\like{\paramvec^{*}}{\dat}}\right]\PDF{\paramvec^*}}{\Kernel{\paramvec^{*}}{\paramvec_{m}}\like{\paramvec_{m}}{\dat}\left[\dfrac{\mclike{\paramvec_{m}}{\dat}}{\like{\paramvec_{m}}{\dat}}\right]\PDF{\paramvec_{m}}}.
	\end{align*}
	Now, define the random variable $Z = \mclike{\dat}{\paramvec}/\like{\dat}{\paramvec}$ with density $\CondPDF{Z}{\paramvec}$ that represents a scaled likelihood estimator. We apply this change of variable and perform some straightforward algebra to obtain a new representation for $\alpha(\paramvec^*,\paramvec_{m})$ that reveals some interesting structure: 
	\begin{align}
		\alpha(\paramvec^{*},\paramvec_{m}) &= \frac{\Kernel{\paramvec_{m}}{\paramvec^*}\like{\dat}{\paramvec^{*}}Z^*\PDF{\paramvec^*}}{\Kernel{\paramvec^{*}}{\paramvec_{m}}\like{\dat}{\paramvec_{m}}Z\PDF{\paramvec_{m}}} \notag \\
		&= \frac{\CondPDF{Z^*}{\paramvec^*}\CondPDF{Z_{m}}{\paramvec_{m}}}{\CondPDF{Z^*}{\paramvec^*}\CondPDF{Z_{m}}{\paramvec_{m}}} \times \frac{\Kernel{\paramvec_{m}}{\paramvec^*}\like{\dat}{\paramvec^{*}}Z^*\PDF{\paramvec^*}}{\Kernel{\paramvec^{*}}{\paramvec_{m}}\like{\dat}{\paramvec_{m}}Z\PDF{\paramvec_{m}}}  \notag \\
		&= \frac{ \CondPDF{Z_{m}}{\paramvec_{m}} \Kernel{\paramvec_{m}}{\paramvec^*}\left[\like{\dat}{\paramvec^{*}}\PDF{\paramvec^*}Z^* \CondPDF{Z^*}{\paramvec^*}\right]}{ \CondPDF{Z^*}{\paramvec^*}\Kernel{\paramvec^{*}}{\paramvec_{m}}\left[\like{\dat}{\paramvec_{m}}\PDF{\paramvec_{m}}Z\CondPDF{Z_{m}}{\paramvec_{m}}\right]}. \label{eqn:step1}
	\end{align}
	The expressions outside the brackets can be considered a proposal density,\\ $\Kernel{Z^*,\paramvec^*}{Z_{m},\paramvec_{m}} = \CondPDF{Z^*}{\paramvec^*}\Kernel{\paramvec^{*}}{\paramvec_{m}}$, in the product state space $\mathcal{Z}\times \paramspace$ where $\mathcal{Z} \subset \mathbb{R}^+$ is the space of values $Z$ can take. By extending the dimension of the Markov chain state by including $Z_{m}$, we see that acceptance probability for the original pseudo-marginal Markov chain in $\paramspace$, as given in \eqref{eqn:step1}, can also be considered as an acceptance probability for this new Markov chain in $\mathcal{Z}\times \paramspace$ based on exact Metropolis-Hastings MCMC. That is,
	\begin{equation*}
		\alpha((Z^*,\paramvec^{*}),(Z_{m},\paramvec_{m})) = \frac{\Kernel{Z_{m},\paramvec_{m}}{Z^*,\paramvec^*}\left[\like{\dat}{\paramvec^{*}}\PDF{\paramvec^*}Z^* \CondPDF{Z^*}{\paramvec^*}\right] }{\Kernel{Z^*,\paramvec^*}{Z_{m},\paramvec_{m}}\left[\like{\dat}{\paramvec_{m}}\PDF{\paramvec_{m}}Z\CondPDF{Z_{m}}{\paramvec_{m}} \right] }.
	\end{equation*}
	This Markov chain has the stationary distribution
	\begin{align*}
		\PDF{Z,\paramvec} &= \like{\dat}{\paramvec}\PDF{\paramvec} Z \CondPDF{Z}{\paramvec}\\
		&\propto \CondPDF{\paramvec}{\dat} Z \CondPDF{Z}{\paramvec}.
	\end{align*}
	Integrating out $Z$ we obtain
	\begin{align*}
		\int_{\mathcal{Z}} \CondPDF{\paramvec}{\dat} Z \CondPDF{Z}{\paramvec} \, \text{d}Z &=  \CondPDF{\paramvec}{\dat} \int_{\mathcal{Z}}  Z \CondPDF{Z}{\paramvec} \, \text{d}Z \\
		&= \CondPDF{\paramvec}{\dat}\CondE{Z}{\paramvec}.
	\end{align*} 
	By linearity of expectation we have
	\begin{align*}
		\CondPDF{\paramvec}{\dat}\CondE{Z}{\paramvec} = \frac{\CondPDF{\paramvec}{\dat}}{\like{\dat}{\paramvec}}\CondE{\mclike{\dat}{\paramvec}}{\paramvec}.
	\end{align*}
	We also have  $\CondE{\mclike{\dat}{\paramvec}}{\paramvec} = \like{\dat}{\paramvec}$, since the Monte Carlo estimator for the likelihood is unbiased. Therefore,
	\begin{equation*}
		\int_{\mathcal{Z}} \PDF{Z,\paramvec} \, \text{d}Z \propto \CondPDF{\paramvec}{\dat}.
	\end{equation*}
	We conclude that the original chain has as its stationary distribution the exact posterior, $\CondPDF{\paramvec}{\dat}$.
	\section{MCMC convergence diagnostics}
	\label{sec:app_conv_diag}
	In the main text we apply the rank normalised $\hat{R}$ statistic and the multiple chain effective sample size measure, $S_\text{eff}$, as defined in the recent work by~\cite{Vehtari2019} that improves earlier definitions~\citep{Gelman1992,Gelman2014}. Since these diagnostics are relatively recent updates, we present their definitions here (see \texttt{Diagnostics.jl} for example implementation).
	
	Consider, $\mathcal{R}$ chains, taking values in $\mathbb{R}^d$, each consisting of an even number of iterations, $\mathcal{M}$. Let $\param_{k,m}^r$ denote the $k$th dimension of the $m$th iteration of the $r$th chain, then we define the rank normalised transform as
	\begin{equation*}
		z_{k,m}^r = \Phi^{-1}\left(\frac{\eta_{k,m}^r - 1/2}{\mathcal{R}\mathcal{M}}\right),
	\end{equation*}
	where $\eta_{k,m}^r$ is the rank of $\param_{k,m}^r$ taken over all $m = 1,2,\ldots, \mathcal{M}$ and $r = 1,2,\ldots,\mathcal{R}$, and $\Phi^{-1} : [0,1] \rightarrow \mathbb{R}$ is the inverse cumulative distribution function of the standard Normal distribution. Then the rank normalised $\hat{R}_k$ statistic for the $k$ parameter is defined as
	\begin{equation*}
		\hat{R}_k = \sqrt{\frac{V_k}{W_k}},
	\end{equation*}
	where
	\begin{equation*}
		V_k = \frac{\mathcal{M}-2}{\mathcal{M}}W_k + \frac{2}{\mathcal{M}}B_k,
	\end{equation*}
	with within-chain variance estimate $W_k$ and between-chain variance estimate $B_k$. These estimates are given by
	\begin{equation*}
		B_k = \frac{\mathcal{M}}{4\mathcal{R} -2} \sum_{r=1}^{\mathcal{R}} \left(\bar{z}_{k,+}^{r} - \bar{\bar{z}}_k\right)^2 +\left(\bar{z}_{k,-}^{r} - \bar{\bar{z}}_k\right)^2\quad \text{and} \quad W_k = \frac{1}{2\mathcal{R}} \sum_{r=1}^\mathcal{R} s_{k,+}^r + s_{k,-}^r,
	\end{equation*} 
	where
	\begin{equation*}
		\begin{split}
			\bar{z}_{k,+}^r = \frac{2}{\mathcal{M}} \sum_{m=\mathcal{M}/2+1}^{\mathcal{M}} z_{k,m}^r,\quad \bar{z}_{k,-}^r = \frac{2}{\mathcal{M}} \sum_{m=1}^{\mathcal{M}/2} z_{k,m}^r,\quad \bar{\bar{z}}_k = \frac{1}{2\mathcal{R}}\sum_{r=1}^{\mathcal{R}} \bar{z}_{k,+}^r + \bar{z}_{k,-}^r\\
			s_{k,+}^r = \frac{2}{\mathcal{M}-2} \sum_{m=\mathcal{M}/2+1}^{\mathcal{M}} \left(z_{k,m}^r  - \bar{z}_{k,+}^r\right)^2\quad\text{and}\quad s_{k,-}^r = \frac{2}{\mathcal{M}-2}\sum_{m=1}^{\mathcal{M}/2} \left(z_{k,m}^r - \bar{z}_{k,-}^r\right)^2.	
		\end{split}
	\end{equation*}
	
	The multiple chain effective sample size measure is computed according to
	\begin{equation*}
		S_{\text{eff},k} = \frac{\mathcal{R}\mathcal{M}}{\hat{\tau}_k},
	\end{equation*} 
	where
	\begin{equation*}
		\hat{\tau}_k = 1 + 2 \sum_{\ell=1}^{L_k}\hat{\rho}_{k,\ell}, \quad \hat{\rho}_{k,\ell} = 1 - \frac{1}{V_k}\left(W_k - \dfrac{1}{\mathcal{R}}\sum_{r=1}^\mathcal{R}\hat{\rho}_{k,\ell}^r\right), 
	\end{equation*}
	and $\hat{\rho}_{k,\ell}^r = \C{z_{k,m}^r}{z_{k,m+\ell}^r}/\V{z_{k,m}^r}$ is the autocorrelation function for the trace of the $k$th dimension of the $r$th chain at lag $\ell$. $L_k$ is the largest odd integer such that $\hat{\rho}_{k,\ell+1} + \hat{\rho}_{k,\ell+2} > 0$ for all $\ell = 1,3,\dots, L_k-2$ \citep{Gelman2014,Vehtari2019}.
	
	To reliably use the chains $\paramvec_{m}^1,\paramvec_{m}^2,\ldots,\paramvec_{m}^\mathcal{R}$ for estimation of the posterior mean, \cite{Vehtari2019} recommend that the chains should at least satisfy the conditions $\hat{R}_k < 1.01$ and $S_{\text{eff},k} > 400$ for all $k = 1,2 ,\ldots, d$. Of course, this does not guarantee that the chains have converged, but it is a guide that, coupled with trace plots and ACF plots, provide reasonably conservative results.
	\section{Observed data}
	\label{sec:obs_data}
	The synthetic data used in the main manuscript and example code is provided in Table~\ref{tab:proddegdat} for the stationary production-degradation model model, Table~\ref{tab:michmentdat} for the Michaelis-Menten model, Table~\ref{tab:schlogldat} for the Schl\"{o}gl model and Table~\ref{tab:reprdat} for the repressilator model.
	
	\begin{table}[h!]
		\centering
		\caption{Data, $\mathcal{D}$,  used for inference on the production-degradation model. Generated using parameter values $k_1 = 1.0$ and $k_2 = 0.01$, initial conditions $X_0 = 10$, and final time $t  = 1,000,000$. }	
		\begin{tabular}{r|rrrrrrrrrr}
			& $Y_{\text{obs}}^{(1)}$ & $Y_{\text{obs}}^{(2)}$ & $Y_{\text{obs}}^{(3)}$ & $Y_{\text{obs}}^{(4)}$ & $Y_{\text{obs}}^{(5)}$ & $Y_{\text{obs}}^{(6)}$ & $Y_{\text{obs}}^{(7)}$ & $Y_{\text{obs}}^{(8)}$ & $Y_{\text{obs}}^{(9)}$ & $Y_{\text{obs}}^{(10)}$ \\
			\hline
			$X_\infty$ &91.68 & 101.64 & 88.13 & 98.88 & 96.36  & 119.59 & 100.62 & 105.11 & 105.30 & 97.00
		\end{tabular}
		\label{tab:proddegdat}
	\end{table}
	\begin{table}[h!]
		\centering
		\caption{Data, $\mathcal{D}$,  used for inference on the Michaelis-Menten model. Generated using parameter values $k_1 = 0.001$, $k_2 = 0.05$ and $k_3 = 0.01$, and initial conditions $E_0 = 100$, $S_0= 100$, $C_0 = 0$ and $P_0$. The observation error is Gaussian with standard deviation $\sigma = 10$.}	
		\begin{tabular}{r|rrrrrrrrrr}
			&$Y_{\text{obs}}^{(1)}$ & $Y_{\text{obs}}^{(2)}$ & $Y_{\text{obs}}^{(3)}$ & $Y_{\text{obs}}^{(4)}$ & $Y_{\text{obs}}^{(5)}$ & $Y_{\text{obs}}^{(6)}$ & $Y_{\text{obs}}^{(7)}$ & $Y_{\text{obs}}^{(8)}$ & $Y_{\text{obs}}^{(9)}$ & $Y_{\text{obs}}^{(10)}$ \\
			\hline
			$t$   & $5$   & $10$  & $15$  & $20$  & $25$  & $30$  & $35$  & $40$  & $45$  & $50$\\
			$E_t$ & 60.84 & 47.21 & 39.53 & 48.64 & 28.99 & 43.53 & 43.78 & 73.16 & 38.40 & 36.84 \\
			$S_t$ & 60.77 & 45.40 & 46.47 & 58.84 & 12.21 & 48.05 & 39.03 & 20.26 & 0.00 & 7.73 \\
			$C_t$ & 42.22 & 62.48 & 54.47 & 59.77 & 60.34 & 61.04 & 57.59 & 67.03 & 50.46 & 64.41  \\
			$P_t$ & 0.00 & 0.00 & 0.00 & 0.00 & 21.60 & 10.04 & 15.78 & 20.71 & 32.32 & 32.34 \\
			\hline
			& & & & & & & & & & \\
			
			&$Y_{\text{obs}}^{(11)}$ & $Y_{\text{obs}}^{(12)}$ & $Y_{\text{obs}}^{(13)}$ & $Y_{\text{obs}}^{(14)}$ & $Y_{\text{obs}}^{(15)}$ & $Y_{\text{obs}}^{(16)}$ & $Y_{\text{obs}}^{(17)}$ & $Y_{\text{obs}}^{(18)}$ & $Y_{\text{obs}}^{(19)}$ & $Y_{\text{obs}}^{(20)}$ \\
			\hline
			$t$   & $55$  & $60$  & $65$  & $70$  & $75$  & $80$  & $85$  & $90$  & $95$ & $100$\\
			$E_t$ & 37.87 & 37.62 & 45.81 & 34.28 & 49.84 & 50.68 & 41.92 & 42.47 & 41.36 & 63.29 \\
			$S_t$ &  1.13 & 15.99 & 17.57 &  5.06 &  5.28 & 0.00 &  4.07 & 17.85 & 19.97 & 27.57\\
			$C_t$ & 64.41 & 49.31 & 53.41 & 62.54 & 55.42 & 42.85 & 43.01 & 62.41 & 37.86 & 38.02 \\
			$P_t$ & 17.16 & 36.85 & 42.20 & 27.55 & 41.33 & 15.40 & 28.60 & 29.29 & 41.10 & 63.48
		\end{tabular}
		\label{tab:michmentdat}
	\end{table}
	\begin{table}[h!]
		\centering
		\caption{Data, $\mathcal{D}$,  used for inference on the Schl\"{o}gl model. Generated using parameter values $k_1 = 0.18$, $k_2 = 0.00025$, $k_3 = 2200.0$ and $k_4 = 37.5$, and initial condition $X_0 = 0$. The observation error is Gaussian with standard deviation $\sigma = 10$.}	
		\begin{tabular}{r|rrrrrrrr}
			&$Y_{\text{obs}}^{(1)}$ & $Y_{\text{obs}}^{(2)}$ & $Y_{\text{obs}}^{(3)}$ & $Y_{\text{obs}}^{(4)}$ & $Y_{\text{obs}}^{(5)}$ & $Y_{\text{obs}}^{(6)}$ & $Y_{\text{obs}}^{(7)}$ & $Y_{\text{obs}}^{(8)}$ \\
			\hline
			$t$ & $12.5$ & $25$ & $37.5$ & $50$ & $62.5$ & $75$ & $87.5$ & $100$ \\
			$X_t$ & 134.99 & 95.83 & 370.91 & 94.15 & 470.12 & 108.17 & 111.20 & 59.54 \\
			\hline
			& & & & & & & & \\
			&$Y_{\text{obs}}^{(9)}$ & $Y_{\text{obs}}^{(10)}$ & $Y_{\text{obs}}^{(11)}$ & $Y_{\text{obs}}^{(12)}$ & $Y_{\text{obs}}^{(13)}$ & $Y_{\text{obs}}^{(14)}$ & $Y_{\text{obs}}^{(15)}$ & $Y_{\text{obs}}^{(16)}$ \\
			\hline
			$t$ & $112.5$ & $125$ & $137.5$ & $150$ & $162.5$ & $175$ & $187.5$ & $200$ \\
			$X_t$ & 99.74 & 347.01 & 92.66 &  377.61 & 416.85 & 120.85 & 361.12 & 282.14
		\end{tabular}
		
		\label{tab:schlogldat}
	\end{table}
	\begin{table}[h!]
		\centering
		\caption{Data, $\mathcal{D}$,  used for inference on the repressilator model. Generated using parameter values $\alpha = 1000$, $\alpha_0 = 1$, $n = 2$ $\beta = 5$ and $\gamma = 1$, and initial conditions $M_{1,0} = 0$, $P_{1,0} = 2$, $M_{2,0} = 0$, $P_{2,0} = 1$, $M_{3,0} = 0$ and $P_{3,0} = 3$. The observation error is Gaussian with standard deviation $\sigma = 10$.}	
		\begin{tabular}{r|rrrrrrrrrr}
			&$Y_{\text{obs}}^{(1)}$ & $Y_{\text{obs}}^{(2)}$ & $Y_{\text{obs}}^{(3)}$ & $Y_{\text{obs}}^{(4)}$ & $Y_{\text{obs}}^{(5)}$ & $Y_{\text{obs}}^{(6)}$ & $Y_{\text{obs}}^{(7)}$ & $Y_{\text{obs}}^{(8)}$ & $Y_{\text{obs}}^{(9)}$ & $Y_{\text{obs}}^{(10)}$ \\
			\hline
			$t$   & $5$   & $10$  & $15$  & $20$  & $25$  & $30$  & $35$  & $40$  & $45$  & $50$\\
			$M_{1,t}$ & 54.09 & 0.00 & 303.69 & 5.46 & 157.54 & 0.00 & 21.27 & 7.88 & 0.00 & 179.26 \\
			$P_{1,t}$ & 45.65 & 6.54 & 416.10 & 17.78 & 141.51 & 26.29 & 4.97 & 24.36 & 21.30 & 264.20 \\
			$M_{2,t}$ & 0.00 & 498.38 & 27.63 & 70.33 & 44.20 & 13.44 & 75.85 & 0.00 & 448.41 & 8.83  \\
			$P_{2,t}$ & 0.00 & 532.66 & 2.56 & 32.25 & 67.96 & 0.00 & 86.08 & 0.00 & 562.18 & 7.73 \\
			$M_{3,t}$ & 20.30 & 11.97 & 27.58 & 213.90 & 0.00 & 376.09 & 6.92 & 439.23 & 0.00 & 323.84 \\
			$P_{3,t}$ & 19.87 & 46.03 & 3.61 & 270.08 & 7.55 & 287.98 & 12.58 & 413.75 & 0.00 & 227.21 \\
			\hline
			& & & & & & & & & & \\
			
			&$Y_{\text{obs}}^{(11)}$ & $Y_{\text{obs}}^{(12)}$ & $Y_{\text{obs}}^{(13)}$ & $Y_{\text{obs}}^{(14)}$ & $Y_{\text{obs}}^{(15)}$ & $Y_{\text{obs}}^{(16)}$ & $Y_{\text{obs}}^{(17)}$ & $Y_{\text{obs}}^{(18)}$ & $Y_{\text{obs}}^{(19)}$ & $Y_{\text{obs}}^{(20)}$ \\
			\hline
			$t$   & $55$  & $60$  & $65$  & $70$  & $75$  & $80$  & $85$  & $90$  & $95$ & $100$\\
			$M_{1,t}$  & 0.00 & 277.77 & 40.45 & 0.00 & 222.82 & 0.00 & 231.91 & 58.95 & 8.22 & 144.86 \\
			$P_{1,t}$ & 0.13 & 177.32 & 24.31 & 0.00 & 241.58 & 19.47 & 157.78 & 70.00 & 6.84 & 120.19 \\
			$M_{2,t}$ & 66.17 & 58.86 & 3.03 & 608.50 & 7.38 & 40.98 & 61.72 & 6.91 & 217.20 & 0  \\
			$P_{2,t}$ & 46.77 & 67.67 & 10.50 & 579.68 & 9.09 & 48.01 & 95.46 & 13.10 & 270.19 & 13.40 \\
			$M_{3,t}$ & 136.19 & 4.53 & 135.78 & 18.84 & 8.62 & 146.78 & 0.00 & 404.93 & 26.80 & 29.69 \\
			$P_{3,t}$ & 197.56  & 3.01 & 136.12 & 8.00 & 5.35 & 174.83 & 0.00 & 393.10 & 22.28 & 8.00 
		\end{tabular}
		\label{tab:reprdat}
	\end{table}
\end{appendices}

\end{document}